\documentclass[prd,aps,showpacs,floats,letterpaper,floatfix,
groupedaddress,superscriptaddress
,nofootinbib
,eqsecnum
]{revtex4}
\usepackage{amsmath}
\usepackage{amsfonts}
\usepackage{bm}
\usepackage{graphicx}
\newcommand{\gothg}{\mathfrak{g}}
\def\nva{{\bf v}_1 \cdot {\bf n}}
\def\nvb{{\bf v}_2 \cdot {\bf n}}
\def\v1v2{{\bf v}_1 \cdot {\bf v}_2}
\allowdisplaybreaks

\begin{document}

\title{Compact binary systems in scalar-tensor gravity: Equations of motion to 2.5 post-Newtonian order}

\author{
Saeed Mirshekari} \email{smirshekari@wustl.edu}
\affiliation{McDonnell Center for the Space Sciences, Department of
Physics, Washington University, St.  Louis, Missouri 63130, USA}
\affiliation{Department of Physics, University of Florida, Gainesville, Florida 32611, USA}

\author{
Clifford M.~Will} \email{cmw@physics.ufl.edu}
\affiliation{McDonnell Center for the Space Sciences, Department of
Physics, Washington University, St.  Louis, Missouri 63130, USA}
\affiliation{Department of Physics, University of Florida, Gainesville, Florida 32611, USA}
\affiliation{GReCO, Institut d'Astrophysique de Paris, CNRS,\\ 
Universit\'e Pierre et Marie Curie, 98 bis Bd. Arago, 75014 Paris, France}

\date{\today}

\begin{abstract}

We calculate the explicit equations of motion for non-spinning compact objects to 2.5 post-Newtonian order, or $O(v/c)^5$ beyond Newtonian gravity, in a general class of scalar-tensor theories of gravity.
We use the formalism of the Direct Integration of the Relaxed Einstein Equations (DIRE), adapted to scalar-tensor theory, coupled with an approach pioneered by Eardley for incorporating the internal gravity of compact, self-gravitating bodies.  For the conservative part of the motion, we obtain the two-body Lagrangian and conserved energy and momentum through second post-Newtonian order.   We find the 1.5 post-Newtonian and 2.5 post-Newtonian contributions to gravitational radiation reaction, the former corresponding to the effects of dipole gravitational radiation, and verify that the resulting energy loss agrees with earlier calculations of the energy flux.  For binary black holes we show that the motion through 2.5 post-Newtonian order is observationally identical to that predicted by general relativity.  For mixed black-hole neutron-star binary systems, the motion is identical to that in general relativity through the first post-Newtonian order, but deviates from general relativity beginning at 1.5 post-Newtonian order, in part through the onset of dipole gravitational radiation.   But through $2.5$ post-Newtonian order, those deviations in the motion of a mixed system are governed by a single parameter dependent only upon the coupling constant $\omega_0$ and the structure of the neutron star, and are formally the same for a general class of scalar-tensor theories as they are for pure Brans-Dicke theory.

\end{abstract}

\maketitle

\section{Introduction and summary}
\label{sec:intro}

The anticipated detection of gravitational waves by a network of ground-based laser-interferometric observatories promises
a new way of ``listening'' to the universe in the high-frequency band. 
A future space-borne interferometer would open the low-frequency band and pulsar timing arrays may soon begin exploring the nano-Hertz region of the gravitational-wave spectrum.  In addition to providing a wealth of astrophysical information, these observations also hold the promise of providing tests of Einstein's theory of general relativity in the strong-field, dynamical regime.

The ``inspiralling compact binary''
-- a binary system of neutron stars or black holes (or one of each) in
the late stages of inspiral and coalescence -- is a leading potential source for detection. 
Given the expected sensitivity of
the
ground-based interferometers, stellar-mass compact binaries could be
detected out to
hundreds of megaparsecs, while for a space interferometer, inspirals involving supermassive
black holes could be heard to cosmological distances.

In order to maximize the detection capability and the science return of these observatories, extremely accurate,
theoretically generated ``templates'' for the gravitational waveform emitted during the inspiral phase must be available.   This means that 
correction
terms in the equations of motion and gravitational-wave signal must be calculated to high orders
in the post-Newtonian (PN) approximation to general relativity,
which, roughly speaking, is an expansion in powers of $v/c \sim (Gm/rc^2)^{1/2} $ (for a review and references see~\cite{sathyabern}).   Contributions to the waveform from the merger phase of the two objects and from the ``ringdown'' phase of the final vibrating black hole also play an important role.

The detected gravitational-wave signals can also be used to test Einstein's theory in the radiative regime, particularly for waves emitted by sources characterized by strong-field gravity, such as inspiraling compact binaries. 
One way to study the potential for this is to check the consistency of a hypothetical observed waveform with the predicted higher-order terms in the general relativistic PN sequence, which depend on very few parameters (only the two masses, for non-spinning, quasi-circular inspirals). 
Another is to examine the constraints that could be placed on specific alternative theories using gravitational-wave observations~\cite{willST,willgraviton,scharrewill,willyunes,BBW,BBW2,stavridis,arunwill09,sopuertayunes}.  Most of these analyses have incorporated only the dominant effect that distinguishes the chosen theory from general relativity, such as dipole radiation or the wavelength-dependent propagation of a massive graviton (see, however~\cite{yunespanicardoso}).   

Some authors have taken a different approach by proposing parametrized versions of the gravitational waveform model~\cite{nicopretorius,mishra,myw,arun12}, inspired by the parametrized post-Newtonian (PPN) formalism used for solar-system experiments, and analysing the bounds that could be placed on those theory-dependent parameters by various gravitational-wave observations.   Yet the authors of these frameworks were limited by the fact that for many alternative theories of gravity, only the leading terms in the waveform model have been derived.  

In addition, the existing parametrizations of the gravitational waveform make the implicit assumption that the gravitational wave signal during the inspiral depends only on the masses of the orbiting compact bodies (in the spinless case), and not on their internal structure.  This is true in general relativity, which satisfies the Strong Equivalence Principle, but is known to be violated by almost every alternative theory that has ever been studied.   In scalar-tensor theory, for example, the internal gravitational binding energy of neutron stars has a definite effect on the motion and gravitational-wave emission, and since the binding energy can amount to as much as 20 percent of the total mass-energy of the body, the effects can be significant.  In order to determine the full nature of the gravitational-wave signal in an alternative theory of gravity, the strong internal gravity of each body must be accounted for somehow, even in a PN expansion.

To make the situation even more interesting, binary black holes play a special role within the scalar-tensor class of alternative theories.   Based on evidence from a 1972 theorem by Hawking~\cite{hawking}, together with known results from first-post-Newtonian theory, it is likely that in a broad class of scalar-tensor theories,  {\em binary black hole motion and gravitational radiation emission are observationally indistinguishable from their GR counterparts}.   This conjecture will be discussed in more detail later in this paper.   

Scalar-tensor gravity is the most popular and well-motivated class of alternative theories to general relativity.   Apart from the long history of such theories, dating back more than 50 years to Jordan, Fierz, Brans and Dicke~\cite{brans}, scalar-tensor gravity has been postulated as a possible low-energy limit of string theory.  In addition, a wide class of so-called $f(R)$ theories, designed to provide an alternative explanation for the acceleration of the universe to the conventional dark-energy model, can be recast into the form of a scalar-tensor theory (for reviews, see~\cite{fujiimaeda,tsujikawa}).  

Measurements in the solar system and in binary pulsar systems already place strong constraints on key parameters of such theories, notably the coupling parameter $\omega_0$.   Yet these tests probe only the lowest-order, first post-Newtonian limit of these theories, some aspects of their strong-field regimes (related to the strong internal gravity of the neutron stars in binary pulsars) and the lowest-order, dipolar aspects of gravitational radiation damping.    

These considerations have motivated us to develop the full equations of motion and gravitational waveform for compact bodies in a class of scalar-tensor theories to a high order in the PN sequence; this is the first in a projected series of papers aiming to treat this problem in full. 

It should be acknowledged that we do not expect any big surprises.   Damour and Esposito-Far\`ese~\cite{DamourEsposito96}  have shown on general grounds that the available constraints on the scalar-tensor coupling constant $\omega_0$ derived from solar-system experiments  imply that scalar-tensor differences from GR will be small to essentially all PN orders, except for certain regions of scalar-tensor theory space where non-linear effects inside neutron stars, called ``spontaneous scalarization'',  can occur (for a recent example, see \cite{barausse}).   
It is therefore unlikely that we will be able to point to a qualitatively new test of scalar-tensor gravity to be performed with gravitational waves.

Nevertheless we expect to provide a complete and consistent waveform model to an order in the PN approximation comparable to the best models from GR.  With this model it will be possible to carry out parameter estimation analyses for gravitational waves from binary inspiral, and to compare the bounds with those from earlier work that either confined attention to the leading dipole term, such as~\cite{BBW}, or assumed extreme mass ratios, such as~\cite{yunespanicardoso}.

We will use a version of the formalism of ``post-Minkowskian'' theory,  which has proven to be very powerful for deriving the equations of motion and gravitational-wave signal to high post-Newtonian orders in GR.   The specific version is known as Direct Integration of the Relaxed Einstein Equations (DIRE),
based on a framework originally developed by
Epstein and Wagoner~\cite{ew}, extended by Will, Wiseman and
Pati~\cite{magnum,opus,patiwill1,patiwill2}, and applied to numerous problems in post-Newtonian gravity~\cite{kww,kidder,willspinorbit,wangwill,zengwill,mitchellwill}.
DIRE is a self-contained approach in which the
Einstein
equations are cast into their ``relaxed'' form of a flat-spacetime wave
equation together with a harmonic gauge condition, and are solved 
formally as a retarded integral over the past null cone of the
field point.  The ``inner'', or near-zone part of this integral
within a sphere of radius $\lambda$,  a
gravitational wavelength, is approximated in a
slow-motion
expansion using standard techniques; the ``outer'' part, extending
over the
radiation zone, is evaluated using a null integration variable.

DIRE is rather easily adapted to scalar-tensory theory, so that the same methods that have been worked out for GR can be applied here.  It is possible that many other theories that generalize the standard action of general relativity in four spacetime dimensions by adding various fields could be cast in a similar form, permitting a systematic study of their predictions for compact binary inspiral beyond the lowest order in the PN approximation.   Indeed another motivation for this work is to lay out a template for possible extensions to other theories of gravity, such as the Einstein-Aether theory~\cite{aether} or TeVeS~\cite{teves}.

Specifically, the theories we address here are described by the action
\begin{equation}
S = (16\pi)^{-1} \int \left [ \phi R - \phi^{-1} \omega(\phi) g^{\alpha\beta} \partial_\alpha \phi \partial_\beta \phi \right ] \sqrt{-g} d^4x + S_m ( \mathfrak{m}, g_{\alpha\beta}) \,,
\label{STaction}
\end{equation}
where $R$ is the Ricci scalar of the spacetime metric $g_{\alpha\beta}$, $\phi$ is the scalar field, of which $\omega$ is a function.  Throughout, we use the so-called ``metric'' or ``Jordan'' representation, in which the matter action $S_m$ involves the matter fields $\mathfrak{m}$ and the metric only; $\phi$ does not couple directly to the matter (see~\cite{DamourEsposito92} for example, for a representation of this class of theories in the so-called ``Einstein'' representation).  We exclude the possibility of a potential or mass for the scalar field.

In order to incorporate the internal gravity of compact, self-gravitating bodies, we adopt an 
approach pioneered by Eardley~\cite{eardley}, based in part on general arguments dating back to Robert Dicke, 
in which one treats the matter energy-momentum tensor as a sum of delta functions located at the position of each body, 
but assumes that the mass of each body is a function $M_A(\phi)$ of the scalar field.  This reflects the fact that the gravitational binding energy of the body is controlled by the value of the gravitational constant, which is directly related to the value of the background scalar field in which the body finds itself.    Consequently, the matter action will have an {\em effective} dependence on $\phi$, and as a result the field equations will depend on the ``sensitivity'' of the mass of each body to variations in the scalar field, holding the total number of baryons fixed.  The sensitivity of body $A$ is defined by
\begin{equation}
s_A \equiv \left ( \frac{d \ln M_A(\phi)}{d \ln \phi} \right ) \,.
\end{equation}
For neutron stars, the sensitivity depends on the mass and equation of state of the star and is typically of order $0.2$; in the weak-field limit, $s_A$ is proportional to the Newtonian self-gravitational energy per unit mass of the body.  From the theorem of Hawking, for stationary black holes, it is known that $s_{\rm BH} = 1/2$.   

This paper reports the results of a calculation of the explicit equations of motion for binary systems of non-spinning compact bodies, through $2.5$PN order, that is, to order $(v/c)^5$ beyond Newtonian theory.   The post-Newtonian corrections at 1PN and 2PN orders are conservative; we obtain from them expressions for the conserved total energy and linear momentum, and obtain the 2-body Lagrangian from which they can be derived.  There are also terms in the equations of motion at $1.5$PN and $2.5$PN orders.  These are gravitational-radiation reaction terms.  Terms at $1.5$PN order do not occur in general relativity, but in scalar-tensor theories with compact bodies, they are the result of the emission of {\em dipole} gravitational radiation.  At $2.5$PN order, one finds the analogue of the general relativistic quadrupole radiation, together with PN correction effects related to monopole and dipole radiation.   

Not surprisingly the expressions for these quantities are complicated, much more so than their counterparts in general relativity.   On the other hand, they depend on a relatively small number of parameters, related to the value of $\omega(\phi)$ far from the system, where $\phi = \phi_0$, along with its derivatives with respect to $\varphi \equiv \phi/\phi_0$, and the sensitivities $s_1$ and $s_2$ of the two bodies, and their derivatives with respect to $\phi$.   
The parameters and their definitions are shown in Table \ref{tab:params}.

At Newtonian order, the ``bare'' gravitational coupling constant $G$ is related to the asymptotic value of the scalar field, but for two-body systems of compact objects, the coupling is given by the combination $G\alpha$, where
\begin{equation}
\alpha = \frac{3 + 2\omega_0}{4 + 2\omega_0} + \frac{(1-2s_1)(1-2s_2)}{4 + 2\omega_0} \,,
\end{equation}
where $\omega_0 = \omega(\phi_0)$.  At $1$PN order there are two body-dependent parameters, $\bar{\gamma}$ and $\bar{\beta}_A$, $A = 1,2$ (see Table \ref{tab:params} for definitions of the parameters).  For non-compact objects, where $s_A \ll 1$,  $\bar{\gamma}= \gamma -1$ and $\bar{\beta}_A = \beta -1$, where $\gamma$ and $\beta$ are precisely the PPN parameters for scalar-tensor theory, as listed, for example in \cite{tegp}.  At $2$PN order, there are two additional parameters $\delta_A$ and $\chi_A$.  Most of the parameters in Table \ref{tab:params} can be related directly to parameters defined in~\cite{DamourEsposito96,DamourEsposito92}.

\begin{table}
\caption{\label{tab:params} Parameters used in the equations of motion}
\begin{ruledtabular}
\begin{tabular}{clcl}
Parameter&Definition&Parameter&Definition\\
\hline
\multicolumn{2}{l}{\bf Scalar-tensor parameters}&\multicolumn{2}{l}{\bf Equation of motion parameters}\\
$G$&$\phi_0^{-1} (4+2\omega_0)/(3+2\omega_0)$&\multicolumn{2}{l}{\bf Newtonian}\\
$\zeta$&$1/(4+2\omega_0)$&$\alpha $&$1 - \zeta + \zeta (1-2s_1)(1- 2s_2) $
\\
$\lambda_1$&$(d\omega/d\varphi)_0 \zeta^2/(1-\zeta)$&\multicolumn{2}{l}{\bf post-Newtonian}\\
$\lambda_2$&$(d^2\omega/d\varphi^2)_0 \zeta^3/(1-\zeta)$&$\bar{\gamma}$ & $-2 \alpha^{-1}\zeta (1-2s_1)(1-2s_2)$
\\
\multicolumn{2}{l}{\bf Sensitivities}&$\bar{\beta}_1 $&$\alpha^{-2} \zeta (1-2s_2)^2 \left ( \lambda_1 (1-2s_1) + 2 \zeta s'_1 \right )$
\\
$s_A$&$[d \ln M_A(\phi)/d \ln \phi]_0$&$\bar{\beta}_2 $&$\alpha^{-2} \zeta (1-2s_1)^2 \left ( \lambda_1 (1-2s_2) + 2 \zeta s'_2 \right )$
\\
$s'_A$&$[d^2 \ln M_A(\phi)/d \ln \phi^2]_0$&\multicolumn{2}{l}{\bf 2nd post-Newtonian}\\
$s''_A$&$[d^3 \ln M_A(\phi)/d \ln \phi^3]_0$&$\bar{\delta}_1$ &$ \alpha^{-2} \zeta (1-\zeta) (1-2s_1)^2$
\\
&&$\bar{\delta}_2$ &$\alpha^{-2} \zeta (1-\zeta) (1-2s_2)^2 $
\\
&&$\bar{\chi}_1 $&$ \alpha^{-3} \zeta (1-2s_2)^3  \left [ (\lambda_2 -4\lambda_1^2 + \zeta \lambda_1 ) (1-2s_1) -6 \zeta \lambda_1 s'_1 + 2 \zeta^2 s''_1 \right ] 
$
\\ 
&&$\bar{\chi}_2 $&$ \alpha^{-3} \zeta (1-2s_1)^3  \left [ (\lambda_2 -4\lambda_1^2 + \zeta \lambda_1 ) (1-2s_2) -6 \zeta \lambda_1 s'_2 + 2 \zeta^2 s''_2 \right ] $

\end{tabular}
\end{ruledtabular}
\end{table}

Here we will quote the bottom-line result: the two-body equation of motion, expressed in relative coordinates, ${\bf x} \equiv {\bf x}_1 - {\bf x}_2$, through $2$PN order.  This equation is ready-to-use, for example in calculating time derivatives of radiative multipole moments in determining the gravitational-wave signal, which will be the subject of the second paper in this series.  The equation has the form
\begin{eqnarray}
\frac{d^2 {\bf x}}{dt^2} &=& -\frac{G\alpha m}{r^2} {\bf n} 
+ \frac{G\alpha m }{ r^2} \bigl[ \, {\bf n} (A_{PN} + A_{2PN}) 
	+ {\dot r}{\bf v} (B_{PN} + B_{2PN} ) \bigr]  
\nonumber \\
&&
+ \frac{8}{5} \eta \frac{(G\alpha m)^2 }{r^3} 
	\bigl[\dot r {\bf n} (A_{1.5PN}+A_{2.5PN})
	- {\bf v}(B_{1.5PN}+B_{2.5PN})\bigr] \,,
\label{eomfinal}
\end{eqnarray}
where  $r \equiv |{\bf x}|$, ${\bf n}
\equiv {\bf x}/r$, $m \equiv m_1 + m_2$, $\eta \equiv m_1m_2/m^2$, 
${\bf v} \equiv {\bf v}_1 - {\bf v}_2$, and $\dot
r = dr/dt$.  
We use units in which $ c = 1 $.  The leading term is Newtonian gravity.
The next group of terms are the conservative terms, of integer PN order, while the final group are
dissipative radiation-reaction terms, of half-odd-integer PN order.  The
coefficients $A$ and $B$ are given explicitly by
\begin{subequations}
\begin{eqnarray}
A_{PN} &=& -(1+3\eta + \bar{\gamma})v^2 + \frac{3}{2}\eta {\dot r}^2 +2(2+\eta + \bar{\gamma} + \bar{\beta}_{+} - \psi \bar{\beta}_{-} ) \frac{G\alpha m}{r} \,,
\nonumber \\
B_{PN} &=&  2(2-\eta + \bar{\gamma}) \,, \label{eomfinalcoeffsPN}
\\
A_{2PN} &=& -\eta(3-4\eta +\bar{\gamma})v^4 
	+ \frac{1}{2}\left [\eta(13-4\eta + 4\bar{\gamma}) - 4 (1-4\eta) \bar{\beta}_+  
	+4 \psi (1-3\eta) \bar{\beta}_{-}  \right ]v^2\frac{G\alpha m}{r}
	-\frac{15}{8}\eta(1-3\eta){\dot r}^4
\nonumber \\
&&
	+\frac{3}{2}\eta(3-4\eta + \bar{\gamma})v^2{\dot r}^2 
	+\left [2 +25\eta+2\eta^2  +2(1+9\eta)\bar{\gamma}
	+ \frac{1}{2} \bar{\gamma}^2 -4\eta (3 \bar{\beta}_{+} - \psi \bar{\beta}_{-})
	  + 2\bar{\delta}_+ + 2\psi \bar{\delta}_{-} \right ]{\dot r}^2\frac{G\alpha m}{r}
\nonumber \\
&&	
	-\biggl [ 9+\frac{87}{4}\eta + (9+8\eta) \bar{\gamma}
	  + \frac{1}{4}(9 -2\eta) \bar{\gamma}^2 
	  + (8+15\eta+4\bar{\gamma}) \bar{\beta}_{+} 
	  - \psi (8+7\eta+4\bar{\gamma})\bar{\beta}_{-} 
	  \nonumber \\
&&	\qquad
        +(1-2\eta)(\bar{\delta}_+ - 2\bar{\chi}_+) 
        + \psi (\bar{\delta}_{-} + 2\bar{\chi}_{-})
	 - 24\eta \frac{\bar{\beta}_1 \bar{\beta}_2}{\bar{\gamma}}
	\biggr ] \left (\frac{G\alpha m}{r} \right )^2 \,,
\nonumber \\
B_{2PN} &=& \frac{1}{2}\eta(15+4\eta + 8\bar{\gamma})v^2
	-\frac{3}{2}\eta(3+2\eta +2\bar{\gamma}){\dot r}^2
\nonumber \\
&&	
	-\frac{1}{2}\left [ 4 +41\eta+8\eta^2  
	  +4(1+7\eta)\bar{\gamma} + \bar{\gamma}^2 
	  -8\eta (2\bar{\beta}_+ - \psi \bar{\beta}_{-} ) + 4\bar{\delta}_+
	  + 4 \psi \bar{\delta}_{-}
	\right ] \frac{G\alpha m}{r} \,,
\label{eomfinalcoeffs2PN}
\\
A_{1.5PN} &=& \frac{5}{2} \zeta {\cal S}_{-}^2\,,
\nonumber \\
B_{1.5PN} &=& \frac{5}{6} \zeta {\cal S}_{-}^2 \,.
\label{eomfinalcoeffs1.5PN}
\end{eqnarray}
\label{eomfinalcoeffs}
\end{subequations}
The expressions for the $2.5$PN coefficients are lengthy and will be displayed in a later section.  Here the subscripts ``$+$'' and ``$-$'' on various parameters denote sums and differences, so that, for a chosen parameter  $\tau_i$ we define
\begin{eqnarray}
\tau_+ &\equiv& \frac{1}{2} ( \tau_1 + \tau_2) \,,
\nonumber \\
\tau_{-} &\equiv& \frac{1}{2} (\tau_1 - \tau_2) \,.
\end{eqnarray}
The quantity ${\cal S}_-$ and its companion ${\cal S}_+$ (used later) are defined by
\begin{eqnarray}
{\cal S}_- &\equiv& - \alpha^{-1/2} (s_1 - s_2) \,,
\nonumber \\
{\cal S}_+ &\equiv&  \alpha^{-1/2} (1-s_1 - s_2) \,,
\end{eqnarray}
(the significance of these definitions of ${\cal S}_\pm$ will become clear in Sec. \ref{sec:genremarks}), and $\psi$ is defined by
\begin{equation}
\psi \equiv \frac{m_1 - m_2}{m_1 + m_2} = \pm \sqrt{1-4\eta} \,.
\end{equation}

Several things are worth noting about these equations (and indeed about all the two-body equations shown later in this paper).  In the general relativistic limit $\omega_0 \to \infty$, or $\zeta \to 0$, the equations (including the $2.5$PN terms) reduce to those of general relativity, as determined by many authors~\cite{damourderuelle,damour300,kopeikin85,GK86,bfp98,futamase01,patiwill2}.   At $1$PN order, the equations agree with the standard scalar-tensor equations, both for weakly self-gravitating bodies in the general class of theories~\cite{nordtvedtST} (shown within the PPN framework in Sec.\ 6.2 and 7.3 of~\cite{tegp}), and for arbitrarily compact bodies in pure Brans-Dicke theory (as displayed in Sec.\ 11.2 of~\cite{tegp}).

Although a number of authors have obtained partial results in scalar-tensor theory at $2$PN order, notably the metric sufficient to study light deflection at $2$PN order~\cite{niST2pn,deng2pn}, and the generic structure of the $2$PN Lagrangian for $N$ compact bodies~\cite{DamourEsposito96}, our explicit formulae for the $2$PN and $2.5$PN contributions to the two-compact-body equations of motion are new.   

The energy loss that results from the $1.5$ PN and $2.5$ PN terms in the equations of motion is in complete agreement with the energy flux calculated to the corresponding order by Damour and Esposito-Far\`ese~\cite{DamourEsposito92}.

The other interesting limit is that in which both bodies are black holes.  Assuming that Hawking's result that $s_{\rm BH} = 1/2$ applies equally for binary black holes as for isolated black holes, we find that the parameters $\bar{\gamma}$, $\bar{\beta}_A$, $\bar{\delta}_A$ and $\bar{\chi}_A$ all vanish, and $\alpha = 1-\zeta = (3+2\omega_0)/(4+2\omega_0)$.  In this case the equations reduce {\em identically} to those of general relativity through $2.5$PN order, with $G\alpha m_A $ replacing of $Gm_A$ for each body.  In other words, if each mass is rescaled by $(4+2\omega_0)/(3+2\omega_0)$, the scalar-tensor equations of motion for binary black holes, including the 2.5PN terms, become {\em identical} to those in general relativity.   Again this applies to all the equations of motion and related quantites (total energy, Lagrangian), whether for the individual bodies or for the relative motion.  Since the masses of bodies in binary systems are measured purely via the Keplerian dynamics of the system, the rescaling is unmeasurable, and therefore, the dynamics of binary black holes in this class of theories is observationally indistinguishable from the dynamics in general relativity.   Assuming, as we believe will be the case, that this is also true for the gravitational wave emission, the conclusion is that gravitational-wave observations of binary black hole systems will be unable to distinguish between these two theories.   

If only one member of the binary system is a black hole, then $\alpha = 1-\zeta$, and $\bar{\gamma} = \bar{\beta}_A = 0$, so that even at $1$PN order, the equations of motion are {\em identical} to those of general relativity, after rescaling each mass.  Only at $1.5$PN order and above do differences between the two theories occur for the mixed binary system, because of the non-vanishing of ${\cal S}_{-}$ in the dipole radiation reaction term, and the non-vanishing of $\bar{\delta}_1$ (if body 1 is the neutron star) in the $2$PN terms.  However, in this case {\em all} the deviations from general relativity depend on a single parameter $Q$, given by
\begin{equation}
Q \equiv \zeta (1-\zeta)^{-1} (1-2s_1)^2 \,,
\end{equation}
where $s_1$ is the sensitivity of the neutron star.  In particular, all reference to the parameters $\lambda_1$ and $\lambda_2$ disappears, and the motion through $2.5$PN order is identical to that predicted by pure Brans-Dicke theory.   If this conclusion holds true for the gravitational-wave emission, then gravitational-wave observations of mixed black-hole neutron-star binaries will be unable to distinguish between Brans-Dicke theory and its generalizations.   The only caveat is that, for a given neutron star, generalized scalar-tensor theories can predict very different values of its un-rescaled mass and its sensitivity from those predicted by pure Brans-Dicke.

The remainder of this paper provides details.  In Sec.\ \ref{sec:relaxed}, we derive the ``relaxed field equations'' and the associated formal equations of motion in scalar-tensor theories, and write down the formal solutions for the gravitational and scalar fields in terms of solutions of the flat spacetime wave equation.  In Sec.\ \ref{sec:nearzonefields} we describe the formal structure of the fields in the near zone, and in Sec.\ \ref{sec:2.5expansion} we obtain formal solutions for the fields through $2.5$PN order in terms of Poisson-like potentials and time derivatives of system multipole moments.  In Sec.\ \ref{sec:matter} we introduce the Eardley approach for characterizing the compact bodies, rewrite all equations in terms of a ``conserved'' density in which the masses of each body are constant, and arrive at the equations of motion expressed in terms of the redefined potentials.  In Sec.\ \ref{sec:2bodyeom} we apply the methods of~\cite{patiwill2} to express the equations of motion explicitly in terms of masses, positions and velocities of each compact body in a two-body system.  We obtain the 2-body Lagrangian, the conserved total energy and linear momentum, the relative effective one-body equations of motion, and the rate of energy loss due to gravitational-radiation reaction.  Section \ref{sec:discussion} presents a detailed discussion of the results.

\section{The relaxed field equations in scalar-tensor theory}
\label{sec:relaxed}

\subsection{Field equations and equations of motion}
\label{sec:fieldequations}

We begin by recasting the field equations of scalar-tensor theory into a form that parallels as closely as possible the ``relaxed Einstein equations'' used to develop post-Minkowskian and post-Newtonian theory in general relativity.    The original field equations of scalar-tensor theory as derived from the action of Eq.\ (\ref{STaction}) take the form
\begin{subequations}
\begin{eqnarray}
G_{\mu\nu} &=& \frac{8\pi}{\phi} T_{\mu\nu} + \frac{\omega(\phi)}{\phi^2} \left ( \phi_{,\mu} \phi_{,\nu} - \frac{1}{2} g_{\mu\nu} \phi_{,\lambda} \phi^{,\lambda} \right ) + \frac{1}{\phi} \left ( \phi_{;\mu\nu} - g_{\mu\nu} \Box_g \phi \right ) \,, 
\label{fieldeq1}\\
\Box_g \phi  &=& \frac{1}{3 + 2\omega(\phi)} \left ( 8\pi T - 16\pi \phi \frac{\partial T}{\partial \phi} - \frac{d\omega}{d\phi} \phi_{,\lambda} \phi^{,\lambda} \right ) \,,
\label{fieldeq2}
\end{eqnarray}
\end{subequations}
where $T_{\mu\nu}$ is the stress-energy tensor of matter and non-gravitational fields, $G_{\mu\nu}$ is the Einstein tensor constructed from the physical metric $g_{\mu\nu}$, $\phi$ is the scalar field, $\omega(\phi)$ is a coupling function, $\Box_g$ denotes the scalar d'Alembertian with respect to  the metric, and commas and semicolons denote ordinary and covariant derivatives, respectively.  We work throughout in the metric or ``Jordan'' representation of the theory, in contrast to the ``Einstein'' representation used, for example in~\cite{DamourEsposito92}.

Normally, such as for a perfect-fluid source, the matter stress-energy tensor depends only on the matter field variables and the physical metric $g_{\mu\nu}$, not on the scalar field, and accordingly the term $\partial T/\partial \phi$ does not appear in the field equations.  But in dealing with a system of self-gravitating bodies, we will adopt an approach pioneered by Eardley~\cite{eardley}.  
Because $\phi$ controls the local value of the gravitational constant in and near each body in this class of theories, the total mass of each body, including its self-gravitational binding energy, may depend on the scalar field.  Thus, as long as each body can be regarded as being in stationary equilibrium during its motion, Eardley proposed letting each mass be a function of $\phi$, namely $M_A(\phi)$.     With this assumption, $T^{\mu\nu}$ takes the form
\begin{eqnarray}
T^{\mu\nu} (x^\alpha) &=& (-g)^{-1/2} \sum_A \int d\tau M_A (\phi) u_A^\mu u_A^\nu \delta^4 (x_A^\alpha (\tau) - x^\alpha)  
\nonumber \\
&=& (-g)^{-1/2} \sum_A M_A (\phi) u_A^\mu u_A^\nu (u_A^0)^{-1} 
\delta^3 ({\bf x} - {\bf x}_A) \,,
\label{Tmunu}
\end{eqnarray}
where $\tau$ is proper time measured along the world line of body $A$ and  $u_A^\mu$ is its four-velocity.  The {\em indirect} coupling of $\phi$ to matter via the binding energy is responsible for the term $\partial T/\partial \phi$ in the field equations.

From the Bianchi identity applied to Eq.\ (\ref{fieldeq1}), the equation of motion is
\begin{equation}
{T^{\mu\nu}}_{;\nu} = \frac{\partial T}{\partial \phi} \phi^{,\mu} \,,
\end{equation}
with the right-hand-side vanishing in the perfect-fluid case.
From the compact body form of $T^{\mu\nu}$ in Eq.\ (\ref{Tmunu}), it can then be shown that the equation of motion for each compact body takes the modified geodesic form
\begin{equation}
u^\nu \nabla_\nu (M_A(\phi) u^\mu ) = - \frac{dM_A}{d\phi} \phi^{,\mu} \,,
\end{equation}
or in terms of coordinate time and ordinary velocities $v^\alpha$,
\begin{equation}
\frac{dv^j}{dt} + \Gamma^j_{\alpha\beta} v^\alpha v^\beta
-\Gamma^0_{\alpha\beta} v^\alpha v^\beta v^j 
= - \frac{1}{M_A(u^0)^2} \frac{dM_A}{d\phi} ( \phi^{,j} - \phi^{,0} v^j ) \,.
\label{geodesiceq}
\end{equation}
These equations of motion could also be derived directly from the effective matter action, $S_m = \sum_A \int_A M_A (\phi) d\tau$.  
Equation (\ref{Tmunu}) can equally well be taken to describe a pressureless perfect fluid (dust), simply by letting the mass of each particle be a constant, independent of $\phi$.

\subsection{Relaxed field equations}
\label{sec:relaxed2}
 
To recast Eq.~(\ref{fieldeq1}) into the form of a ``relaxed'' Einstein equation, we make use of the following well-known property:  defining the 
quantities 
\begin{subequations}
\begin{eqnarray}
\gothg^{\mu\nu} &\equiv& \sqrt{-g} g^{\mu\nu} \,, 
\label{gothg}\\
H^{\mu\alpha\nu\beta} &\equiv& \gothg^{\mu\nu}  \gothg^{\alpha\beta} -  \gothg^{\alpha\nu} \gothg^{\beta\mu} \,,
\label{Hdef}
\end{eqnarray}
\end{subequations}
 it can be shown that the following is an identity, valid for any spacetime,
\begin{equation}
{H^{\mu\alpha\nu\beta}}_{,\alpha\beta} = (-g) (2G^{\mu\nu} + 16\pi t_{LL}^{\mu\nu} ) \,,
\label{Hidentity}
\end{equation}
where $t_{LL}^{\mu\nu}$ is the Landau-Lifshitz pseudotensor [see Eq.\ (20.22) of \cite{mtw} for an explicit formula].

To incorporate scalar-tensor theory into this framework, we assume that, far from any isolated source, the metric takes its Minkowski form $\eta_{\mu\nu}$, and that the scalar field $\phi$ tends to a constant value $\phi_0$.   We define the rescaled scalar field $\varphi \equiv \phi/\phi_0$.  We next define the conformally transformed metric $\tilde{g}_{\mu\nu}$ by 
\begin{equation}
\tilde{g}_{\mu\nu} \equiv \varphi {g}_{\mu\nu} \,,
\label{gtilde}
\end{equation}
and the gravitational field $\tilde{h}^{\mu\nu}$ by the equation
\begin{equation}
\tilde{\gothg}^{\mu\nu} \equiv \sqrt{-\tilde{g}} \tilde{g}^{\mu\nu} \equiv \eta^{\mu\nu} - \tilde{h}^{\mu\nu} \,.
\label{hdef}
\end{equation}
From Eq.\ (\ref{gtilde}) it can be shown that this is equivalent to
\begin{equation}
\gothg^{\mu\nu} \equiv \varphi^{-1} (\eta^{\mu\nu} - \tilde{h}^{\mu\nu}) \,.
\end{equation}
We now impose the ``Lorentz'' gauge condition
\begin{equation}
{\tilde{h}^{\mu\nu}}_{,\nu} = 0 \,,
\label{gauge}
\end{equation}
which is equivalent to
\begin{equation}
{\gothg^{\mu\nu}}_{,\nu} =- \varphi^{-2} \varphi_{,\nu} (\eta^{\mu\nu}- \tilde{h}^{\mu\nu} ) \,.
\end{equation}
Substituting Eqs.\ (\ref{fieldeq1}), (\ref{fieldeq2}), (\ref{hdef}) and (\ref{gauge}) into (\ref{Hidentity}), we can recast the field equation (\ref{fieldeq1}) into the form
\begin{equation}
\Box_\eta \tilde{h}^{\mu\nu} =  -16\pi \tau^{\mu\nu} \,,
\label{heq0}
\end{equation}
where $\Box_\eta$ is the flat spacetime d'Alembertian with respect to $\eta_{\mu\nu}$, and where
\begin{equation}
16\pi \tau^{\mu\nu}= 16\pi (-g) \frac{\varphi}{\phi_0} T^{\mu\nu}  + \Lambda^{\mu\nu} + \Lambda_S^{\mu\nu} \,,
\label{heq}
\end{equation}
where
\begin{subequations}
\begin{eqnarray}
\Lambda^{\mu\nu} &\equiv& 16\pi \left [ (-g) t_{LL}^{\mu\nu} \right ](\tilde{\gothg}^{\mu\nu})
+ {\tilde{h}^{\mu\alpha}}_{,\beta} {\tilde{h}^{\nu\beta}}_{,\alpha}
-\tilde{h}^{\alpha\beta} {\tilde{h}^{\mu\nu}}_{,\alpha\beta} \,,
\\
\Lambda_S^{\mu\nu} &\equiv&  \frac{(3+ 2\omega)}{\varphi^2} \varphi_{,\alpha} \varphi_{,\beta} \left ( \tilde{\gothg}^{\mu\alpha}\tilde{\gothg}^{\nu\beta} - \frac{1}{2} \tilde{\gothg}^{\mu\nu}\tilde{\gothg}^{\alpha\beta} \right ) \,,
\end{eqnarray}
\label{Lambdadef}
\end{subequations}
where the notation $[(-g) t_{LL}^{\mu\nu}](\tilde{\gothg}^{\mu\nu})
$ denotes that the Landau-Lifshitz piece should be calculated using only $\tilde{\gothg}$, in other words, exactly as in general relativity, except using the conformal metric, rather than the physical metric.   
The scalar field equation can also be rewritten in terms of a flat-spacetime wave equation, of the form
\begin{equation}
\Box_\eta \varphi = -8\pi \tau_s \,,
\label{phieq}
\end{equation}
where
\begin{eqnarray}
 \tau_s &=& -\frac{1}{3+2\omega} \sqrt{-g} \frac{\varphi}{\phi_0} \left ( T -2 \varphi \frac{\partial T}{\partial \varphi} \right )
- \frac{1}{8\pi}  \tilde{h}^{\alpha\beta} \varphi_{,\alpha\beta}
\nonumber \\
&&
+\frac{1}{16\pi} \frac{d}{d\varphi} \left [ \ln \left ( \frac{3+2\omega}{\varphi^2} \right ) \right ] \varphi_{,\alpha} \varphi_{,\beta} \tilde{\gothg}^{\alpha\beta} \,.
\label{taustar}
\end{eqnarray}
In principle, Eqs.\ (\ref{gothg}) and (\ref{hdef}) can be combined to give $g_{\mu\nu}$ in terms of $\varphi$ and $\tilde{h}^{\mu\nu}$, although in practice, we will express it as a PN expansion.  The final result will be the relaxed field equations (\ref{heq0}) - (\ref{taustar}) expressed entirely in terms of $\tilde{h}^{\mu\nu}$, $\varphi$, and the matter variables.   The next task will be to solve these equations iteratively in a post-Newtonian expansion in the near-zone.
Formally the solutions of these wave equations can be expressed using the standard retarded Green function, in the form
\begin{eqnarray}
\tilde{h}^{\mu\nu} (t,{\bf x}) &=& 4 \int \frac{\tau^{\mu\nu} (t-|{\bf x}-{\bf x}'|,{\bf x}')}{|{\bf x}-{\bf x}'|} d^3x' \,,
\nonumber \\
\varphi (t,{\bf x}) &=& 2 \int \frac{\tau_s (t-|{\bf x}-{\bf x}'|,{\bf x}')}{|{\bf x}-{\bf x}'|} d^3x' \,,
\end{eqnarray}
where the integration is over the past flat spacetime null cone of the field point $(t,{\bf x})$.   We will expand these integrals in the near-zone, and incorporate a slow-motion, weak-field expansion in terms of a small parameter $\epsilon \sim v^2 \sim m/r$; the strong-field internal gravity effects will be encoded in the functions $M_A(\phi)$.

\section{Formal structure of the near-zone fields}
\label{sec:nearzonefields}

We follow \cite{patiwill1} (hereafter referred to as PWI) by defining a simplified notation for the field
$\tilde{h}^{\mu\nu}$ and the scalar field $\varphi$:
\begin{eqnarray}
N &\equiv& \tilde{h}^{00} \sim O(\epsilon) \,, \nonumber \\
K^i &\equiv& \tilde{h}^{0i} \sim O(\epsilon^{3/2}) \,, \nonumber \\
B^{ij} &\equiv& \tilde{h}^{ij} \sim O(\epsilon^2) \,, \nonumber \\
B &\equiv& \tilde{h}^{ii} \equiv \sum_i \tilde{h}^{ii} \sim O(\epsilon^2) \,,
\nonumber \\
\Psi &\equiv& \varphi - 1 \sim O(\epsilon) \,,
\end{eqnarray}
where we show the leading order dependence on $\epsilon$ in the near
zone.  To obtain
the equations of motion to 2.5PN order, we need to determine
the components of the physical metric and $\varphi$ to the following orders:
$g_{00}$ to $O(\epsilon^{7/2})$,
$g_{0i}$ to $O(\epsilon^{3})$ ,
$g_{ij}$ to $O(\epsilon^{5/2})$, and 
$\varphi$ to $O(\epsilon^{7/2})$.
From the definitions (\ref{gothg}) and  (\ref{hdef}), one can invert to find
$g_{\mu\nu}$ in terms of $\tilde{h}^{\mu\nu}$ and $\varphi$ to the appropriate order in $\epsilon$, as in PWI, Eq.\ (4.2).   
Expanding to the
required order, we find,
\begin{subequations}
\label{metricexpand}
\begin{eqnarray}
g_{00} &=& -1 +  \left ( \frac{1}{2} N + \Psi \right) 
+  \left ( \frac{1}{2} B - \frac{3}{8} N^2 - \frac{1}{2} N \Psi
- \Psi^2 \right )
\nonumber \\
&& +  \left (\frac{5}{16} N^3 - \frac{1}{4} NB + \frac{1}{2} K^j K^j +\frac{3}{8} N^2 \Psi - \frac{1}{2} B \Psi +\frac{1}{2} N \Psi^2 + \Psi^3 \right )
\nonumber \\
&&  +O(\epsilon^4) \,, 
\label{metric00}
\\
g_{0i} &=& -  K^i  +  \left ( \frac{1}{2} N + \Psi \right ) K^i  +O(\epsilon^{7/2}) \,, 
\label{metric0i}
\\
g_{ij} &=& \delta^{ij} \left \{ 1+  \left (\frac{1}{2} N - \Psi \right ) -  \left ( \frac{1}{8} N^2 + \frac{1}{2} B + \frac{1}{2} N \Psi - \Psi^2 \right ) \right \} +  B^{ij} 
\nonumber \\
&& +O(\epsilon^3) \,, 
\label{metricij}
\\
(-g) &=& 1+  (N - 4\Psi) -  (B +4N \Psi - 10 \Psi^2 ) + O(\epsilon^3) \,.
\label{metricdet}
\end{eqnarray}
\label{metric}
\end{subequations}
In Eqs.\ (\ref{metricexpand}) we do not distinguish between covariant and contravariant components of quantities such as $K^i$ or $B^{ij}$, since their indices are assumed to be raised or lowered using the Minkowski metric, whose spatial components are $\delta_{ij}$.

We now define a set of provisional ``densities'' following the convention of Blanchet and Damour~\cite{blanchetdamour}, but adding a separate density for the scalar field equation:\begin{eqnarray}
\sigma &\equiv& T^{00} + T^{ii} \,,
\nonumber \\
\sigma^i &\equiv& T^{0i} \,,
\nonumber \\
\sigma^{ij} &\equiv& T^{ij} \,.
\nonumber \\
\sigma_s &\equiv& - T + 2\varphi \partial T/\partial \varphi \,.
\label{sigmas}
\end{eqnarray}
The second contribution to $\sigma_s$ will be non-zero only in the case where our system consists of gravitationally bound bodies, whose internal structure could depend on the environmental value of $\varphi$.

Because of the way we have formulated the relaxed scalar-tensor equations, the quantity $\Lambda^{\mu\nu}$ has {\em exactly} the same form as in PWI, Eq.\ (4.4).  To the 2PN order needed for our work, we have
\begin{subequations}
\label{Lambda}
\begin{eqnarray}
\Lambda^{00} &=& - \frac{7}{8} (\nabla N)^2 + \left \{ \frac{5}{8} {\dot N}^2 - \ddot N N -2 {\dot N}^{,k} K^k
+ \frac{1}{2} K^{i,j} (3 K^{j,i}+ K^{i,j}) \right . 
\nonumber \\
&& \left . 
+ {\dot K}^j N^{,j} - B^{ij} N^{,ij} + 
\frac{1}{4} \nabla N \cdot
\nabla B + \frac{7}{8} N (\nabla N)^2 \right \}  + O(\rho \epsilon^3) \,,
\\
\Lambda^{0i} &=& \left \{ N^{,k}( K^{k,i}- K^{i,k}) + \frac{3}{4}\dot
N N^{,i} \right \} + O(\rho \epsilon^{5/2}) \,, \\
\Lambda^{ij} &=& \frac{1}{4} \{ N^{,i}N^{,j} - \frac{1}{2}
\delta^{ij} (\nabla N)^2 \} 
+ \left \{ 2 K^{k,(i}K^{j),k}- K^{k,i}K^{k,j}
-K^{i,k}K^{j,k} +2N^{,(i} {\dot K}^{j)} + \frac{1}{2}N^{,(i} B^{,j)} \right . \nonumber \\
&& \left . - \frac{1}{2}N(N^{,i}N^{,j} - \frac{1}{2}
\delta^{ij} (\nabla N)^2) -\delta^{ij} (K^{l,k}K^{[k,l]} 
+N^{,k}{\dot K}^{k}
+ \frac{3}{8}{\dot N}^2 + \frac{1}{4} \nabla N \cdot
\nabla B ) \right \} 
 + O(\rho \epsilon^3) \,,
\\
\Lambda^{ii} &=& - \frac{1}{8} (\nabla N)^2 
+ \left \{ K^{l,k}K^{[k,l]}-N^{,k}{ \dot K}^{k} - \frac{1}{4}\nabla N \cdot \nabla B - \frac{9}{8} {\dot N}^2
+ \frac{1}{4}N ({\nabla N})^2 \right \} 
 +  O(\rho \epsilon^3) \,.
\end{eqnarray}
\end{subequations}
To the required order, the scalar stress-energy pseudotensor is given by
\begin{subequations}
\label{Lambdascalar}
\begin{eqnarray}
\Lambda_S^{00} &=& \frac{3+2\omega_0}{2} (\nabla \Psi )^2
+\frac{3+2\omega_0}{2} \left \{ N (\nabla \Psi )^2 - 2 \left ( 1- \frac{\omega_0'}{3+2\omega_0} \right ) \Psi  (\nabla \Psi )^2 + {\dot \Psi}^2 \right \} 
\nonumber \\
&& + O(\rho \epsilon^3) \,,
\\
\Lambda_S^{0i} &=& -(3+2\omega_0) {\dot \Psi} \Psi^{,i} +  O(\rho \epsilon^{5/2}) \,, 
\\
\Lambda_S^{ij} &=&  (3+2\omega_0) \left \{ \Psi^{,i}\Psi^{,j} - \frac{1}{2}
\delta^{ij} (\nabla \Psi)^2 \right \} \nonumber \\
&&-(3+2\omega_0) \left \{  2\left ( 1- \frac{\omega_0'}{3+2\omega_0} \right ) \Psi 
\left [\Psi^{,i}\Psi^{,j} - \frac{1}{2}\delta^{ij} (\nabla \Psi)^2 \right ] -\frac{1}{2} \delta^{ij} {\dot \Psi}^2
\right \} 
 + O(\rho \epsilon^3) \,,
\\
\Lambda_S^{ii} &=& -\frac{3+2\omega_0}{2}(\nabla \Psi)^2 
+ (3+2\omega_0) \left \{ \left ( 1- \frac{\omega_0'}{3+2\omega_0} \right ) \Psi (\nabla \Psi)^2 + \frac{3}{2} {\dot \Psi}^2 \right \}
 +  O(\rho \epsilon^3) \,,
\end{eqnarray}
\end{subequations}
where $\omega'_0 \equiv (d\omega/d\varphi)_0$.

The near-zone expansions of the fields $N$, $K^i$, $B^{ij}$ and $\Psi$ are then given by
\begin{subequations}
\label{bigexpansion}
\begin{eqnarray}
%
%
N_{\cal N} &=& 4 \epsilon \int_{\cal M} \frac{\tau^{00}(t,{\bf x}^\prime)}{|{\bf x}-{\bf
x}^\prime |} d^3x^\prime 
+2 \epsilon^2 \partial^2_t \int_{\cal M} \tau^{00}(t,{\bf x}^\prime) |{\bf x}-{\bf x}^\prime | d^3x^\prime
-\frac{2}{3} \epsilon^{5/2} \stackrel{(3)\qquad}{{\cal I}^{kk}(t)} 
+ \frac{1}{6} \epsilon^3 \partial^4_t \int_{\cal M}
\tau^{00}(t,{\bf x}^\prime) |{\bf x}-{\bf x}^\prime |^3 d^3x^\prime 
\nonumber \\
&&- \frac{1}{30} \epsilon^{7/2} \left \{ (4x^{kl}+2r^2\delta^{kl})
\stackrel{(5)\quad}{{\cal I}^{kl}(t)}
- 4 x^k \stackrel{(5)\qquad}{{\cal I}^{kll}(t)}
+ \stackrel{(5)\qquad}{{\cal I}^{kkll}(t)} \right
\} 
+ N_{\partial {\cal M}} + O(\epsilon^4) \,,
\label{bigexpansiona} \\
%
%
K^i_{\cal N} &=& 4 \epsilon^{3/2} \int_{\cal M} \frac{\tau^{0i}(t,{\bf x}^\prime)}{|{\bf x}-{\bf
x}^\prime |} d^3x^\prime +2 \epsilon^{5/2} \partial^2_t \int_{\cal M}
\tau^{0i}(t,{\bf x}^\prime) |{\bf x}-{\bf x}^\prime | d^3x^\prime
 +\frac{2}{9} \epsilon^3 \left \{ 3 x^k \stackrel{(4)\quad}{{\cal I}^{ik}(t)}
- \stackrel{(4)\qquad}{{\cal I}^{ikk}(t)}
+2 \epsilon^{mik}  \stackrel{(3)\qquad}{{\cal J}^{mk}(t)}
\right \} \nonumber \\
&&+ K^i_{\partial {\cal M}} + O(\epsilon^{7/2}) \,,
\label{bigexpansionb} \\
%
%
B^{ij}_{\cal N} &=& 4 \epsilon^2 \int_{\cal M} \frac{\tau^{ij}(t,{\bf x}^\prime)}{|{\bf x}-{\bf x}^\prime |} d^3x^\prime 
- 2 \epsilon^{5/2} \stackrel{(3)\quad}{{\cal I}^{ij}(t)}
+2 \epsilon^3 \partial^2_t \int_{\cal M}
\tau^{ij}(t,{\bf x}^\prime) |{\bf x}-{\bf x}^\prime | d^3x^\prime
\nonumber \\
&& - \frac{1}{9} \epsilon^{7/2} \left \{ 
3 r^2 \stackrel{(5)\quad}{{\cal I}^{ij}(t)}
-2x^k \stackrel{(5)\qquad}{{\cal I}^{ijk}(t)}
- 8 x^k \epsilon^{mk(i} \stackrel{(4)\qquad}{{\cal J}^{m|j)}(t)}
+ 6 \stackrel{(3)\qquad}{M^{ijkk}(t)} \right \}
 + B^{ij}_{\partial {\cal M}} + O(\epsilon^4) \,,
\label{bigexpansionc}\\
%
%
\Psi_{\cal N} &=& 2 \epsilon \int_{\cal M} \frac{\tau_s(t,{\bf x}^\prime)}{|{\bf x}-{\bf x}^\prime |} d^3x^\prime 
- 2 \epsilon^{3/2} \dot{M_s}  
+ \epsilon^2 \partial^2_t \int_{\cal M} \tau_s(t,{\bf x}^\prime) |{\bf x}-{\bf x}^\prime | d^3x^\prime
\nonumber 
\\
&&
-\frac{1}{3} \epsilon^{5/2} \left ( r^2 \stackrel{(3)\quad}{M_s(t)}
-2x^j \stackrel{(3)\quad}{{\cal I}^{j}_s(t)}
+ \stackrel{(3)\quad}{{\cal I}^{kk}_s(t)} \right )
+ \frac{1}{12} \epsilon^3 \partial^4_t \int_{\cal M}
\tau_s(t,{\bf x}^\prime) |{\bf x}-{\bf x}^\prime |^3 d^3x^\prime 
\nonumber \\
&&- \frac{1}{60} \epsilon^{7/2} \left \{ r^4 \stackrel{(5)\quad}{M_s(t)} 
- 4r^2 x^j \stackrel{(5)\quad}{{\cal I}^{j}_s(t)}
+(4x^{kl}+2r^2\delta^{kl})
\stackrel{(5)\quad}{{\cal I}^{kl}_s(t)}
- 4 x^k \stackrel{(5)\qquad}{{\cal I}^{kll}_s(t)}
+ \stackrel{(5)\qquad}{{\cal I}^{kkll}_s(t)} \right
\} 
 + O(\epsilon^4) \,,
 \label{bigexpansiond}
\end{eqnarray}
\end{subequations}
where we define the moments of the system by
\begin{subequations}
\begin{eqnarray}
{\cal I}^Q &\equiv&  \int_{\cal M} \tau^{00} x^Q d^3x \,,
\label{IQ}
\\
{\cal J}^{iQ} &\equiv&  \epsilon^{iab}\int_{\cal M} \tau^{0b} x^{aQ} d^3x \,,
\label{JiQ}
\\
M^{ijQ} &\equiv& \int_{\cal M} \tau^{ij} x^Q d^3x \,,
\label{MijQ}
\\
{\cal I}^{Q}_s &\equiv& \int_{\cal M} \tau_s x^Q d^3x \,,
\label{IsQ}
\\
M_s &\equiv& \int_{\cal M} \tau_s d^3x \,.
\label{Ms}
\end{eqnarray}
\label{genmoment}
\end{subequations}
The index $Q$ is a
multi-index, such that $x^Q$ denotes $x^{i_1} \dots x^{i_q}$.
The integrals are taken over a constant time hypersurface $\cal M$ at time $t$ out to a radius $\cal R$, which represents the boundary between the near zone and the far zone.
The structure of the expansions for ${N}_{\cal N}$, 
${K}^i_{\cal N}$ and ${B}^{ij}_{\cal N}$ is identical to the structure in PWI because the source $\tau^{\mu\nu}$ satisfies the conservation law ${\tau^{\mu\nu}}_{,\nu}=0$, a consequence of the Lorentz gauge condition.    However, no such explicit conservation law applies to $\tau_s$; nevertheless, in a post-Newtonian expansion, we will be able to show, for example, that the term $\epsilon^{3/2} \dot{M}_s$ actually vanishes to lowest PN order, and thus contributes only beginning at $\epsilon^{5/2}$ order; the other terms involving time derivatives of $M_s$ will also be boosted to one higher PN order.  The time derivatives of the dipole moments ${\cal I}^{j}_s$ do {\em not} vanish in general; this is related to the well-known phenomenon of dipole gravitational radiation that can occur in scalar-tensor theories.
The boundary terms $N_{\partial {\cal M}}$, $K^i_{\partial {\cal M}}$ and $B^{ij}_{\partial {\cal M}}$ can be found in Appendix C of PWI, but they will play no role in our analysis.  As in PWI, we will discard all terms that depend on the radius $\cal R$ of the near-zone; these necessarily cancel against terms that arise from integrating over the remainder of the past null cone; those ``outer'' integrals can be shown to make no contribution to the near zone metric to the PN order at which we are working.

In the near zone, the potentials are Poisson-like potentials and their generalizations.  Most were defined in \cite{patiwill1}, but we will need to define additional potentials associated with the scalar field.  For a source
$f$, we define the Poisson potential to be
\begin{equation}
P(f) \equiv \frac{1}{4\pi} \int_{\cal M} \frac{f(t,{\bf x}^\prime)}
{|{\bf x}-{\bf x}^\prime | } d^3x^\prime \,, \quad \nabla^2
P(f) = -f \,.
\end{equation}
We also define potentials based on the ``densities'' $\sigma$,
$\sigma^i$ and $\sigma^{ij}$ and $\sigma_s$ constructed from $T^{\alpha\beta}$ and from $T-2\varphi \partial T/\partial \varphi$,
\begin{subequations}
\begin{eqnarray}
\Sigma (f) &\equiv& \int_{\cal M} \frac{\sigma(t,{\bf x}^\prime)f(t,{\bf x}^\prime)}{|{\bf x}-{\bf x}^\prime | } d^3x^\prime = P(4\pi\sigma f) \,,
\\
\Sigma^i (f) &\equiv& \int_{\cal M} \frac{\sigma^i(t,{\bf x}^\prime)f(t,{\bf
x}^\prime)}{|{\bf x}-{\bf x}^\prime | } d^3x^\prime = P(4\pi\sigma^i f) \,,
\\
\Sigma^{ij} (f) &\equiv& \int_{\cal M} \frac{\sigma^{ij}(t,{\bf x}^\prime)f(t,{\bf x}^\prime)}{|{\bf x}-{\bf x}^\prime | } d^3x^\prime = P(4\pi\sigma^{ij} f) \,,
\\
\Sigma_s (f) &\equiv& \int_{\cal M} \frac{\sigma_s(t,{\bf x}^\prime)f(t,{\bf
x}^\prime)}{|{\bf x}-{\bf x}^\prime | } d^3x^\prime = P(4\pi\sigma_s f) \,,
\end{eqnarray}
\end{subequations}
along with the superpotentials
\begin{subequations}
\label{definesuper}
\begin{eqnarray}
X(f)  &\equiv& \int_{\cal M} {\sigma(t,{\bf x}^\prime)f(t,{\bf
x}^\prime)}
{|{\bf x}-{\bf x}^\prime | } d^3x^\prime  \,,
\\
Y(f) &\equiv& \int_{\cal M} {\sigma(t,{\bf x}^\prime)f(t,{\bf x}^\prime)}
{|{\bf x}-{\bf x}^\prime |^3 } d^3x^\prime  \,,
\end{eqnarray}
\end{subequations}
and their obvious counterparts $X^i$,  $X_s$,  and so on.
A number of potentials occur sufficiently frequently in the PN
expansion that it is useful to define them specifically.  There are the ``Newtonian'' potentials,
\begin{subequations}
\begin{eqnarray}
U \equiv \int_{\cal M} \frac{\sigma(t,{\bf x}^\prime)}{|{\bf x}-{\bf x}^\prime | } d^3x^\prime = P(4\pi\sigma) =
\Sigma(1) \,,
\\
U_s\equiv \int_{\cal M} \frac{\sigma_s (t,{\bf x}^\prime)}{|{\bf x}-{\bf x}^\prime | } d^3x^\prime = P(4\pi\sigma_s) =
\Sigma_s(1) \,.
\end{eqnarray}
\end{subequations}
The potentials needed for the post-Newtonian limit are:
\begin{eqnarray}
V^i &\equiv& \Sigma^i(1) \,, \quad \Phi_1^{ij} \equiv \Sigma^{ij}(1) \,, \quad
 \Phi_1 \equiv \Sigma^{ii}(1) \,, \quad \Phi_1^s \equiv \Sigma_s (v^2) \,,\nonumber \\
\Phi_2 &\equiv& \Sigma(U) \,, \quad \Phi^s_2 \equiv \Sigma_s(U) \,,
\quad \Phi_{2s} \equiv \Sigma (U_s) \,, \quad \Phi^s_{2s} \equiv \Sigma_s (U_s) \,,
\nonumber \\
X &\equiv& X(1)\,, \quad X_s \equiv X_s(1) \,.
\end{eqnarray}
Useful 2PN potentials include:
\begin{eqnarray}
&V_2^i \equiv \Sigma^i(U) \,, \qquad & V_{2s}^i \equiv \Sigma^i(U_s) \,,
\nonumber \\
&\Phi_2^i \equiv \Sigma(V^i) \,, \qquad & Y \equiv Y(1) \,,
\nonumber \\
&X^i \equiv X^i(1) \,, \qquad & X_1 \equiv X^{ii}(1) \,, \nonumber
\\
&X_2 \equiv  X(U) \,, \qquad & X_{2s} \equiv X(U_s) \,,\nonumber \\
&X_2^s \equiv X_s(U) \,,\qquad & X_{2s}^s \equiv X_s(U_s) \,,\nonumber \\
&P_2^{ij} \equiv P(U^{,i}U^{,j}) \,, \qquad & P_2 \equiv P_2^{ii}=\Phi_2
-\frac{1}{2}U^2 \,,\nonumber \\
&P_{2s}^{ij} \equiv P(U^{,i}_sU^{,j}_s) \,, \qquad & P_{2s} \equiv P_{2s}^{ii}=\Phi_{2s}^s
-\frac{1}{2}U^{2}_s \,,\nonumber \\
&G_1 \equiv P({\dot U}^2)  \,, \qquad & G_{1s} \equiv P({\dot U}_s^{2}) \,,
\nonumber \\
&G_2 \equiv P(U {\ddot U})  \,, \qquad & G_{2s} \equiv P(U {\ddot U}_s) \,,
\nonumber \\
&G_3 \equiv -P({\dot U}^{,k} V^k) \,, \qquad & G_{3s} \equiv
 -P({\dot U}_s^{,k} V^k) \,,\nonumber \\
&G_4 \equiv P(V^{i,j}V^{j,i}) \,, \qquad & G_5 \equiv -P({\dot V}^k U^{,k}) \,,\nonumber \\
&G_6 \equiv P(U^{,ij} \Phi_1^{ij}) \,, \qquad & G_{6s} \equiv P(U_s^{,ij} \Phi_1^{ij}) \,,\nonumber \\
&G_7^i \equiv P(U^{,k}V^{k,i}) + \frac{3}{4} P(U^{,i}\dot U ) \,, \nonumber \\
& H \equiv P(U^{,ij} P_2^{ij}) \,, \qquad & H_s \equiv P(U^{,ij} P_{2s}^{ij}) \,, \nonumber \\
& H^s \equiv P(U_s^{,ij} P_2^{ij}) \,, \qquad & H_s^s \equiv P(U_s^{,ij} P_{2s}^{ij}) \,. 
\label{potentiallist}
\end{eqnarray}

\section{Expansion of near-zone fields to 2.5PN order}
\label{sec:2.5expansion}

In evaluating the contributions at each order, we shall use the
following notation,
\begin{subequations}
\label{expandNKB}
\begin{eqnarray}
N &=& \epsilon (N_0 + \epsilon N_1+ \epsilon^{3/2} N_{1.5}+ \epsilon^2 N_2+
\epsilon^{5/2} N_{2.5}  )
+O(\epsilon^4) \,, \\
K^i &=& \epsilon^{3/2} (K_1^i + \epsilon K_2^i +\epsilon^{3/2} K_{2.5}^i ) +O(\epsilon^{7/2}) \,, \\
B &=& \epsilon^2 (B_1 + \epsilon^{1/2} B_{1.5} 
+\epsilon B_2 + \epsilon^{3/2} B_{2.5} )
+O(\epsilon^4) \,, \\
B^{ij} &=& \epsilon^2 (B_2^{ij} + \epsilon^{1/2} B_{2.5}^{ij} 
) +O(\epsilon^3) \,,
\\
\Psi &=& \epsilon (\Psi_0 + \epsilon^{1/2} \Psi_{0.5} +\epsilon \Psi_1+ \epsilon^{3/2} \Psi_{1.5}+ \epsilon^2 \Psi_2+
\epsilon^{5/2} \Psi_{2.5})
+O(\epsilon^4) \,,
\end{eqnarray}
\end{subequations}
where the subscript on each term indicates the level (1PN, 2PN, 2.5PN,
etc.) of its leading contribution to the equations of motion.

\subsection{Newtonian, 1PN and 1.5PN solutions}
\label{sec:N1.5PNsolution}

At lowest order in the PN expansion, we only need to evaluate
$\tau^{00} = (-g)T^{00}(\varphi/\phi_0) + O(\rho\epsilon) = \sigma/\phi_0  +
O(\rho\epsilon)$ (recall that $\sigma^{ii} \sim \epsilon \sigma$), and $\tau_s = \sigma_s/[\phi_0(3+2\omega_0)] +O(\rho\epsilon)$, where $\omega_0 \equiv \omega(\phi_0)$.
Since both densities have compact support, the outer integrals vanish, and we
find
\begin{eqnarray}
N_0 &=& \frac{4U}{\phi_0} \,,
\\
\Psi_0 &=& \frac{2U_s}{\phi_0 (3+2\omega_0)}  \,.
\label{newtonian}
\end{eqnarray}
Consider the case where we are dealing with pure perfect fluids, with no compact bodies having sensitivity factors $s_A$. Then to Newtonian order, $\sigma = \sigma_s$, $U=U_s$, and the metric to Newtonian order is given by the leading term in Eq.\ (\ref{metric00}), 
\begin{eqnarray}
g_{00} &=& -1 + \left ( \frac{1}{2} N + \Psi \right) 
\\
&=& -1 + 2 \frac{4+2\omega_0}{\phi_0 (3+2\omega_0)} U  \,.
\end{eqnarray}
We therefore identify the coefficient of $U$ in $g_{00}$ as the effective Newtonian gravitational coupling constant, $G$, given by  
\begin{equation}
G \equiv \frac{1}{\phi_0} \frac{4+2\omega_0}{3+2\omega_0} \,.
\label{Gdefinition}
\end{equation}
However, we will not set $G=1$ as is conventional in general relativity, in order to highlight the fact that it is an effective gravitational constant linked to the asymptotic value of $\phi$, which could, for example, vary with time as the universe evolves. 
For future use, we also define the parameters
\begin{eqnarray}
\zeta &\equiv&  \frac{1}{4+2\omega_0}
\,, \nonumber 
\\
\lambda_1 &\equiv&  \frac{(d\omega/d\varphi)_0 \zeta}{3+2\omega_0} \,,
\nonumber 
\\
\lambda_2 &\equiv& \frac{(d^2\omega/d\varphi^2)_0 \zeta^2}{3+2\omega_0} \,.
\end{eqnarray}
A consequence of these definitions is that 
\begin{eqnarray}
\frac{1}{\phi_0} &=& G (1-\zeta) \,,
\nonumber \\
\frac{1}{\phi_0 (3+2\omega_0)} &=& G \zeta \,.
\end{eqnarray}
It is worth pointing out that $\omega_0$ enters at Newtonian order, via the modified coupling constant $G$ of Eq.\ (\ref{Gdefinition}).  It is then clear, by virtue of the expansion $\omega(\phi) = \omega_0 + (d\omega/d\varphi)_0 \Psi +
 (d^2\omega/d\varphi^2)_0 \Psi^2/2 + \dots$, that the parameter $\lambda_1$ will first contribute at $1$PN order, $\lambda_2$ will first contribute at $2$PN order, and so on.
 
To this order, $(-g)= 1+ 4GU(1-\zeta)-8GU_s \zeta + O(\epsilon^2)$.
Then, through PN order, the required forms for $\tau^{\mu\nu}$ and $\tau_s$ are given by
\begin{subequations}
\begin{eqnarray}
\tau^{00} &=&G(1-\zeta) \biggl \{  \sigma - \sigma^{ii} +G(1-\zeta) \bigl ( 4\sigma U
- \frac{7}{8\pi} (\nabla U)^2 \bigr ) - G\zeta \bigl ( 6\sigma U_s - \frac{1}{8\pi}  ({\nabla U_s})^{2} \bigr ) \biggr \}
+ O(\rho\epsilon^2) \,, 
\label{tau00PN}
 \\
\tau^{0i} &=& G(1-\zeta) \sigma^i + O(\rho\epsilon^{3/2}) \,,
\label{tau0jPN}
 \\
\tau^{ii} &=& G(1-\zeta) \biggl \{  \sigma^{ii} - \frac{1}{8\pi} G(1-\zeta) (\nabla U)^2 - \frac{1}{8\pi} G\zeta (\nabla U_s)^{2} \biggr \}
+ O(\rho\epsilon^2) \,, 
\label{tauiiPN}
 \\
\tau^{ij} &=&  O(\rho\epsilon) \,, 
\label{tauijPN}
 \\
\tau_s &=& G \zeta \biggl \{ \sigma_s + 2G (1-\zeta) \sigma_s U
 - 2G ( 2\lambda_1 +\zeta ) \sigma_s U_s
 + \frac{1}{2\pi} G( \lambda_1 -\zeta) (\nabla U_s)^{2}   \biggr \} + O(\rho \epsilon^2) 
 \,. 
\label{tausPN}
\end{eqnarray}
\label{tauPN}
\end{subequations}
Substituting into Eqs. (\ref{bigexpansion}), and calculating terms
through 1.5PN order (e.g. $O(\epsilon^{5/2})$ in $N$), we obtain
\begin{subequations}
\label{postnewtonian}
\begin{eqnarray}
N_1 &=& G(1-\zeta) \biggl \{ 7G (1-\zeta) U^2 -4 \Phi_1
+2G (1-\zeta) \Phi_2+2 {\ddot X} 
\nonumber \\
&&\quad - G \zeta  U_s^{2} -24G  \zeta \Phi_{2s}
+ 2 G \zeta  \Phi^s_{2s} \biggr \}
 \,, \\
K_{1}^i &=& 4G(1-\zeta) V^i \,,\\
B_1 &=& G(1-\zeta) \biggl \{ G(1-\zeta)U^2+4\Phi_1-2G(1-\zeta)\Phi_2
\nonumber \\
&& \quad + G\zeta U_s^{2} - 2G\zeta \Phi^s_{2s} \biggr \} \,, \\
\Psi_1 &=& G\zeta \biggl \{ -2G ( \lambda_1-\zeta ) U_s^{2} 
+ 4G (1-\zeta)\Phi^s_2 -4 G( \lambda_1+ 2\zeta ) \Phi^s_{2s}
+  \ddot{X}_s \biggr \} \,,
\\
N_{1.5} &=& -\frac{2}{3} \stackrel{(3)\qquad}{{\cal I}^{kk}(t)}
\,,\\
B_{1.5} &=& -{2} \stackrel{(3)\qquad}{{\cal I}^{kk}(t)} \,,
\\
\Psi_{1.5} &=& -2 \dot{M}_s(t) 
+ \frac{2}{3} x^j \stackrel{(3)\quad}{{\cal I}_s^{j}(t)}
- \frac{1}{3} \stackrel{(3)\qquad}{{\cal I}_s^{kk}(t)} \,.
\label{psi15}
\end{eqnarray}
\end{subequations}
In  Eq.\ (\ref{psi15}), we have used the fact (to be verified later) that, because of the conservation of baryon number, and assuming that our compact bodies have  stationary internal structure, $M_s(t)$ is constant to the lowest PN order.  Thus, rather than contributing to $\Psi_{0.5}$ as shown in Eq.\ (\ref{bigexpansiond}), the term $-2\dot{M}_s$ contributes to $\Psi_{1.5}$; similarly the term in $\Psi_{1.5}$ involving three time derivatives of $M_s$ actually contributes to $\Psi_{2.5}$.

The physical metric to 1.5PN order is then given by
\begin{subequations}
\label{1.5pnmetric}
\begin{eqnarray}
g_{00} &=& -1 + 2G(1-\zeta)U + 2G\zeta U_s  - 2G^2 (1-\zeta)^2 U^2 
- 2G^2 \zeta ( \zeta + \lambda_1) U_s^{2}
\nonumber \\
&& 
-4G^2 \zeta (1-\zeta) U\,U_s
+4G^2 \zeta (1-\zeta) \Phi^s_2
-12 G^2 \zeta (1-\zeta) \Phi_{2s}
\nonumber \\
&&-4G^2 \zeta (2 \zeta + \lambda_1) \Phi^s_{2s}
+ G(1-\zeta) \ddot X 
+G\zeta {\ddot X}_s
\nonumber \\
&&
- \frac{4}{3}\stackrel{(3)\qquad}{{\cal I}^{kk}(t)} 
-2 \dot{M}_s(t) 
+ \frac{2}{3} x^j \stackrel{(3)\quad}{{\cal I}_s^{j}(t)}
- \frac{1}{3} \stackrel{(3)\qquad}{{\cal I}_s^{kk}(t)}
+ O(\epsilon^3) \,,\\
g_{0i} &=& -4G(1-\zeta)V^i + O(\epsilon^{5/2}) \,,\\
g_{ij} &=& \delta_{ij} \bigl [1+2G(1-\zeta)U -2G\zeta U_s \bigr ] + O(\epsilon^2) \,.
\end{eqnarray}
\end{subequations}

\subsection{2PN and 2.5PN solutions}
\label{sec:22.5PNsolutions}

At 2PN and 2.5PN order, we obtain, from Eqs.\ (\ref{heq}), (\ref{Lambda}) and (\ref{Lambdascalar}),
\begin{subequations}
\begin{eqnarray}
\tau^{ij} &=& G(1-\zeta) \sigma^{ij} 
+ \frac{1}{4\pi} G^2 (1-\zeta)^2 \bigl (U^{,i}U^{,j} - \frac{1}{2}
\delta^{ij} (\nabla U)^2 \bigr ) 
\nonumber \\
&& + \frac{1}{4\pi} G^2 \zeta (1-\zeta) \bigl (U_s^{,i}U_s^{,j} - \frac{1}{2}
\delta^{ij} (\nabla U_s)^2 \bigr ) 
+ O(\rho\epsilon^2) \,, \\
\tau^{0i} &=& G(1-\zeta) \sigma^i + G^2 (1-\zeta)^2 \bigl (  4\sigma^i U
+ \frac{2}{\pi} U^{,j}V^{[j,i]}
+ \frac{3}{4\pi} \dot U U^{,i} \bigr )
\nonumber \\
&& 
- G^2 \zeta (1-\zeta) \bigl ( 6\sigma^i U_s
+ \frac{1}{4 \pi}  \dot{U}_s U_s^{,i} \bigr )
+ O(\rho\epsilon^{5/2}) \,.
\end{eqnarray}
\end{subequations}
Outer integrals and boundary terms contribute nothing, so we obtain
\begin{subequations}
\begin{eqnarray}
B_2^{ij} &=& 4G(1-\zeta) \Phi_1^{ij} 
+ G^2 (1-\zeta)^2 \bigl [ 4P_2^{ij}-\delta^{ij}(2\Phi_2-U^2) \bigr ]
+G^2 \zeta (1-\zeta) \bigl [ 4{P}_{2s}^{ij}-\delta^{ij}(2\Phi^s_{2s}-U_s^{2}) \bigr ] \,,
\\
K_2^i &=& G^2 (1-\zeta)^2 \bigl ( 8V_2^i -8\Phi_2^i + 8UV^i + 16G_7^i \bigr ) + 2G(1-\zeta){\ddot X}^i
-G^2 \zeta (1-\zeta) \bigl ( 24 V^i_{2s} +4 P(\dot{U}_s U_s^{,i} ) \bigr ) \,,
\\
B_{2.5}^{ij} &=&
-2 \stackrel{(3)\qquad}{{\cal I}^{ij}(t)}  \,, \\
K_{2.5}^i &=& \frac{2}{3} x^k \stackrel{(4)\qquad}{{\cal
I}^{ik}(t)}
- \frac{2}{9} \stackrel{(4)\qquad}{{\cal I}^{ikk}(t)}
+ \frac{4}{9} \epsilon^{mik}  \stackrel{(3)\qquad}{{\cal J}^{mk}(t)}
\,.
\end{eqnarray}
\end{subequations}
All solutions obtained so far must be substituted into Eqs.\ (\ref{heq}), (\ref{taustar}), (\ref{Lambda}) and (\ref{Lambdascalar}) to obtain $\tau^{00}$, $\tau^{ii}$ and $\tau_s$ to the required order,
\begin{subequations}
\begin{eqnarray}
\tau^{00} &=& G(1-\zeta) \biggl \{ \sigma - \sigma^{ii} 
+G(1-\zeta) \bigl ( 4\sigma U- \frac{7}{8\pi} (\nabla U)^2 \bigr ) 
- G\zeta \bigl ( 6\sigma U_s - \frac{1}{8\pi}  (\nabla U_s)^{2} \bigr ) \biggr \} 
\nonumber \\
&&
+ G^2 (1-\zeta)^2 \biggl \{ \sigma \biggl [ 7G(1-\zeta) U^2 - 8\Phi_1 + 2G (1-\zeta) \Phi_2 + 2\ddot{X} \biggr ]  - 4\sigma^{ii} U 
\nonumber \\
&& \quad + \frac{1}{4\pi} \biggl [ \frac{5}{2} \dot{U}^2 - 4U \ddot{U}
- 8 \dot{U}^{,k} V^k  +2V^{i,j}(3V^{j,i}+V^{i,j}) +4{\dot V}^jU^{,j}
-4U^{,ij}\Phi_1^{ij}
+8\nabla U \cdot \nabla \Phi_1
\nonumber \\
&& \quad
- \frac{7}{2}  \nabla U \cdot \nabla \ddot X
- G(1-\zeta) \left ( 4 \nabla U \cdot \nabla \Phi_2 
+10U(\nabla U)^2 
+4U^{,ij} P_2^{ij} \right ) \biggr ] \biggr \}
\nonumber \\
&&
+ G^2 \zeta (1-\zeta) \biggl \{ \sigma \biggl [
G(6\lambda_1 - 1 + 19\zeta) U_s^{2}
-G(1-\zeta) \bigl ( 24 UU_s
+24 \Phi_{2s}
+12  \Phi_2^s \bigr )
\nonumber \\
&& \quad
+2 G( 6\lambda_1+1+ 11\zeta) \Phi_{2s}^s
-3 \ddot{X}_s \biggr ] + 6\sigma^{ii} U_s
\nonumber \\
&& \quad
+\frac{1}{4\pi} \biggl [ 
 G(1-\zeta) \left ( 2U (\nabla U_s)^2 
+4 U_s \nabla U \cdot \nabla U_s
+42  \nabla U \cdot \nabla \Phi_{2s}
+2 \nabla U_s \cdot \nabla \Phi_{2}^s
-4 \nabla U \cdot \nabla \Phi_{2s}^s
-4 U^{,ij} P_{2s}^{ij}   \right )
\nonumber \\
&& \quad
+\frac{1}{2} \dot{U}_s^{2} 
-2 G(\lambda_1 + 2\zeta)  \nabla U_s \cdot \nabla \Phi_{2s}^s
+\frac{1}{2} \nabla U_s \cdot \nabla \ddot{X}_s
 \biggr ] \biggr \}
 \nonumber \\
&& 
+ G(1-\zeta) \biggl \{ \sigma \biggl [
 \frac{4}{3}   \stackrel{(3)\qquad}{{\cal I}^{kk}(t)} + 6\dot{M}_s(t)
 -2 x^j  \stackrel{(3)\quad}{{\cal I}_s^{j}(t)}
 +  \stackrel{(3)\quad}{{\cal I}_s^{kk}(t)} \biggr ]
+ \frac{1}{2\pi} U^{,ij}   \stackrel{(3)\quad}{{\cal I}^{ij}(t)}
+ \frac{1}{12\pi} U_s^{,j}  \stackrel{(3)\quad}{{\cal I}_s^{j}(t)} \biggr \} 
 \nonumber \\
&& +O(\rho \epsilon^3) \,,
\\
\tau^{ii} &=& G(1-\zeta) \biggl \{ \sigma^{ii}  
- \frac{1}{8\pi} G(1-\zeta)(\nabla U)^2 
-\frac{1}{8\pi} G\zeta (\nabla U_s)^{2} \biggr \}
 \nonumber \\
&&
+ G^2 (1-\zeta)^2 \biggl \{ 4 \sigma^{ii} U 
- \frac{1}{4\pi} \biggl [ \frac{9}{2}{\dot U}^2 
+4V^{i,j}V^{[i,j]} +4{\dot V}^jU^{,j} 
+ \frac{1}{2} \nabla U \cdot \nabla \ddot X \biggr ] \biggr \}
 \nonumber \\
&&
- G^2 \zeta(1-\zeta) \biggl \{ 6 \sigma^{ii} U_s 
- \frac{1}{4\pi} \biggl [ \frac{3}{2} \dot{U}_s^{2} 
-G(1-\zeta) \bigl ( 2\nabla U_s \cdot \nabla \Phi_2^s 
-6 \nabla U \cdot \nabla \Phi_{2s} \bigr )
 \nonumber \\
&& \quad
+2G (\lambda_1 + 2\zeta) \nabla U_s \cdot \nabla \Phi_{2s}^s 
-\frac{1}{2}  \nabla U_s \cdot \nabla \ddot{X}_s \biggr ] \biggr \}
\nonumber \\
&&
-\frac{1}{12\pi} G(1-\zeta)U_s^{,j}  \stackrel{(3)\quad}{{\cal I}_s^{j}(t)}  +O(\rho \epsilon^3)\,, 
\\
\tau_s &=& G\zeta \biggl \{ \sigma_s + 2G(1-\zeta) \sigma_s U
-2G (2\lambda_1 + \zeta) \sigma_s U_s 
+ \frac{1}{2\pi} G (\lambda_1 - \zeta) (\nabla U_s)^{2} \biggr \}
\nonumber \\
&&
+ G^2 \zeta \sigma_s \biggl \{ 
G(1-\zeta) \biggl [ 2(1-\zeta) U^2 
- 4(2\lambda_1+\zeta) \bigl (UU_s + \Phi_2^s \bigr )
-12 \zeta \Phi_{2s}  \biggr ] 
-(1-\zeta) \bigl (  4\Phi_1 - \ddot{X} \bigr )
\nonumber \\
&& \quad
+ G (20 \lambda_1^2 - 4 \lambda_2 + 6 \zeta \lambda_1 +2\zeta^2 ) U_s^{2}
+4G (2\lambda_1 + \zeta)(\lambda_1 + 2\zeta) \Phi_{2s}^s
- (2\lambda_1 + \zeta) \ddot{X}_s \biggr \}
\nonumber \\
&&
- \frac{1}{8\pi} G^2 \zeta \biggl \{
(1-\zeta) \bigl ( 8U \ddot{U}_s + 16 V^j \dot{U}_s^{,j} + 8 \Phi_1^{ij} U_s^{,ij} \bigr ) 
+ 4(\lambda_1 - \zeta) \bigl ( \dot{U}_s^{2} -  \nabla U_s \cdot \nabla \ddot{X}_s \bigr )
\nonumber \\
&& \quad
- G(1-\zeta) \biggl [ 16 (\lambda_1-\zeta) \nabla U_s \cdot \nabla \Phi_{2}^s
- 8(1-\zeta) U_s^{,ij} P_2^{ij} - 8\zeta U_s^{,ij} P_{2s}^{ij} \biggr ]
\nonumber \\
&& \quad
+ 16 G (\lambda_1 +2\zeta)(\lambda_1 - \zeta)\nabla U_s \cdot \nabla \Phi_{2s}^s 
-8 G (\lambda_2 -4\lambda_1^2 + 4\zeta \lambda_1 - \zeta^2 ) U_s (\nabla U_s)^{2} \biggr \} 
\nonumber \\
&&
+ G\biggl \{ \sigma_s \left [
 \frac{2}{3}  \zeta \stackrel{(3)\qquad}{{\cal I}^{kk}(t)} + \frac{1}{3} (2\lambda_1 + \zeta) \left (6\dot{M}_s(t) 
 -2 x^j  \stackrel{(3)\quad}{{\cal I}_s^{j}(t)}
 +  \stackrel{(3)\qquad}{{\cal I}_s^{kk}(t)} \right ) \right ]
 \nonumber \\
&& \quad
+ \frac{1}{2\pi} \zeta U_s^{,ij}   \stackrel{(3)\quad}{{\cal I}^{ij}(t)}
+ \frac{1}{3\pi} (\lambda_1 - \zeta) U_s^{,j}  \stackrel{(3)\quad}{{\cal I}_s^{j}(t)} \biggr \} 
+ O(\rho \epsilon^3) \,.
\end{eqnarray}
\end{subequations}
Substituting into Eqs.\ (\ref{bigexpansiona}), (\ref{bigexpansionc}) and (\ref{bigexpansiond}) 
and evaluating terms
through $O(\epsilon^{7/2})$, and verifying that the outer integrals and
surface terms make no ${\cal R}$-independent contributions, we obtain,
\begin{subequations}
\begin{eqnarray}
N_2 &=& G(1-\zeta) \biggl \{ \frac{1}{6} \stackrel{(4)}{Y} 
- 2 \ddot{X}_1
+ G(1-\zeta) \biggl [ 7U\ddot{X} -16 U\Phi_1 -4 V^i V^i 
-16 \Sigma(\Phi_1) + \Sigma(\ddot{X}) + 8 \Sigma^i (V^i) 
+ \ddot{X}_2
 \nonumber \\
&& \quad
 - 4G_1 - 16G_2 + 32 G_3 + 24 G_4 
- 16 G_5 - 16 G_6 \biggr ]
+G^2(1-\zeta)^2 \biggl [  8U\Phi_2 + \frac{20}{3} U^3 -16 H \biggr ]
\biggr \}
 \nonumber \\
&&
+ G^2 \zeta (1-\zeta) \biggl \{
24 \Sigma^{ii} (U_s) -U_s \ddot{X}_s -12 \Sigma(\ddot{X}_s)
+ \Sigma_s (\ddot{X}_s) -12 \ddot{X}_{2s} + \ddot{X}_{2s}^s
+4 G_{1s}  
 \nonumber \\
&& \quad
+ G(1-\zeta) \biggl [ 8U \Phi_{2s}^s - 4 U_s \Phi_2^s - 84 U \Phi_{2s} - 
4 UU_s^{2} -12\Sigma(\Phi_{2s}) -48 \Sigma(\Phi_2^s)
 \nonumber \\
&& \quad
+4 \Sigma_s(\Phi_2^s) +4\Sigma_s(UU_s) -12 \Sigma(UU_s) -16H_s
\biggr ] + 24 G(\lambda_1 + 3\zeta) \Sigma(U_s^{2}) \nonumber \\
&& \quad
+ 4G(\lambda_1 + 2\zeta)\biggl [ 12 \Sigma(\Phi_{2s}^s)
- \Sigma_s(\Phi_{2s}^s) - \Sigma_s(U_s^{2}) +  U_s \Phi_{2s}^s
 \biggr ]
\biggr \}  \,,
\\
B_2 &=& G(1-\zeta) \biggl \{ 2\ddot{X}_1 
+ G(1-\zeta) \biggl [ U\ddot{X} + 4 V^iV^i - \Sigma(\ddot{X})
-8 \Sigma^i(V^i) +16 \Sigma^{ii}(U) 
 \nonumber \\
&& \quad
 - \ddot{X}_2 - 20G_1
+ 8G_4
+16 G_5 \biggr ] \biggr \}
+ G^2 \zeta(1-\zeta) \biggl \{ U_s \ddot{X}_s -24 \Sigma^{ii} (U_s) 
-\Sigma_s (\ddot{X}_s)  -\ddot{X}_{2s}^s
+4 G_{1s}
 \nonumber \\
&& \quad
+ G(1-\zeta) \biggl [ 4 U_s \Phi_2^s - 12 U \Phi_{2s} 
 +12\Sigma(\Phi_{2s}) 
-4 \Sigma_s(\Phi_2^s) -4\Sigma_s(UU_s) +12 \Sigma(UU_s) 
\biggr ]
 \nonumber \\
&& \quad
+ 4G (\lambda_1 + 2\zeta)\biggl [ 
 \Sigma_s(\Phi_{2s}^s) + \Sigma_s(U_s^{2}) -  U_s \Phi_{2s}^s
 \biggr ] \biggr \} \,,
 \\
\Psi_2 &=&
G \zeta \biggl \{ \frac{1}{12} \stackrel{(4)}{Y_s} + G(1-\zeta) \biggl [ 2 \Sigma_s(\ddot{X}) 
-8 \Sigma_s(\Phi_1)
-8 G_{2s} + 16 G_{3s} - 8G_{6s} + 2\ddot{X}_2^s \biggr ]
  \nonumber \\
&& \quad
 -2 G(\lambda_1 +2\zeta) \bigl ( \Sigma_s(\ddot{X}_s) 
+ \ddot{X}_{2s}^s \bigr )
-2G  (\lambda_1-\zeta) U_s\ddot{X}_s
- 8G^2 (1-\zeta)(\lambda_1 +2\zeta) \biggl [\Sigma_s (\Phi_2^s)
+ \Sigma_s(UU_s) \biggr ]
 \nonumber \\
&& \quad
+ 8 G^2 (\lambda_1 +2\zeta) \biggl [
 (\lambda_1-\zeta) U_s \Phi_{2s}^s
 +(\lambda_1 +2\zeta) \Sigma_s(\Phi_{2s}^s)
  \biggr ] -8 G^2  (1-\zeta)(\lambda_1-\zeta) U_s \Phi_2^s 
  \nonumber \\
&& \quad
+G^2 (1-\zeta)^2 \bigl ( 4\Sigma_s(U^2) -8H^s \bigr )
- G^2 \zeta (1-\zeta)  \bigl ( 24 \Sigma_s(\Phi_{2s}) + 8H_s^s \bigr )
 \nonumber \\
&& \quad
- \frac{4}{3} G^2 ( \lambda_2 -4\lambda_1^2 + 4\zeta \lambda_1 - \zeta^2 ) U_s^{3}
-4 G^2 (\lambda_2-4\lambda_1^2 -5\zeta\lambda_1-4\zeta^2) \Sigma_s(U_s^{2})  \biggr \} \,,
\\
N_{2.5} &=& 
 -\frac{1}{15}(2x^{kl}+r^2\delta^{kl})\stackrel{(5)\quad}{{\cal
I}^{kl}(t)} + \frac{2}{15} x^k\stackrel{(5)\qquad}{{\cal I}^{kll}(t)}
- \frac{1}{30} \stackrel{(5)\qquad}{{\cal I}^{kkll}(t)}
+ G(1-\zeta) \biggl [ \frac{16}{3} U\stackrel{(3)\qquad}{{\cal I}^{kk}(t)}
-4X^{,kl}\stackrel{(3)\qquad}{{\cal I}^{kl}(t)} 
\nonumber \\
&&
\qquad +24 U \dot{M}_s(t)
-8(x^k U - X^{,k})\stackrel{(3)\quad}{{\cal I}_s^{k}(t)}
+4U \stackrel{(3)\qquad}{{\cal I}_s^{kk}(t)}
-\frac{2}{3} X_s^{,k}\stackrel{(3)\quad}{{\cal I}_s^{k}(t)} \biggr ]
\,, \\
B_{2.5} &=&  
-\frac{1}{3}  r^2 \stackrel{(5)\quad}{{\cal I}^{kk}(t)}
+\frac{2}{9} x^k \stackrel{(5)\qquad}{{\cal I}^{kll}(t)}
+ \frac{8}{9}  x^k \epsilon^{mkj} \stackrel{(4)\qquad}{{\cal J}^{mj}(t)}
- \frac{2}{3}  \stackrel{(3)\qquad}{M^{kkll}(t)}
+\frac{2}{3}G(1-\zeta) X_s^{,k}\stackrel{(3)\quad}{{\cal I}_s^{k}(t)} \,,
\\
\Psi_{2.5} &=&
-\frac{1}{30}(2x^{kl}+r^2\delta^{kl})\stackrel{(5)\quad}{{\cal
I}_s^{kl}(t)} 
+ \frac{1}{15} x^k\stackrel{(5)\qquad}{{\cal I}_s^{kll}(t)}
- \frac{1}{60} \stackrel{(5)\qquad}{{\cal I}_s^{kkll}(t)}
+ \frac{1}{15} r^2 x^k\stackrel{(5)\quad}{{\cal I}_s^{k}(t)}
-\frac{1}{3} r^2 \stackrel{(3)\quad}{M_s(t)}
\nonumber \\
&&
+ G\zeta \biggl [ \frac{4}{3} U_s \stackrel{(3)\qquad}{{\cal I}^{kk}(t)}
-2 X_s^{,kl} \stackrel{(3)\quad}{{\cal I}^{kl}(t)} \biggr ]
+ \frac{4}{3} G(\lambda_1+2\zeta) X_s^{,k} \stackrel{(3)\quad}{{\cal I}_s^{k}(t)}
\nonumber \\
&&
+\frac{2}{3}G(2\lambda_1 +\zeta) U_s \biggl [ 6\dot{M}_s(t) 
- 2x^k\stackrel{(5)\quad}{{\cal I}_s^{k}(t)}
+  \stackrel{(3)\quad}{{\cal I}_s^{kk}(t)} \biggr ] \,.
\end{eqnarray}
\end{subequations}

\section{PN expansion of the matter source}
\label{sec:matter}

\subsection{Energy momentum tensor and the conserved density}
\label{sec:conserveddensity}

We now must expand the effective energy-momentum tensor, Eq.\ (\ref{Tmunu}) in a PN expansion to the required order, including the $\phi$ dependence of the masses $M_A$.  
We first expand $M_A (\phi)$ about the asymptotic value $\phi_0$:  
\begin{equation}
M_A(\phi) = M_{A0} + \delta\phi \left (\frac{dM_A}{d\phi} \right )_0
+ \frac{1}{2} \delta \phi^2  \left (\frac{d^2M_A}{d\phi^2} \right )_0
+ \frac{1}{6} \delta \phi^3  \left (\frac{d^3M_A}{d\phi^3} \right )_0
+ \dots \,.
\end{equation}  
We then define the dimensionless ``sensitivities''
\begin{eqnarray}
s_A &\equiv& \left ( \frac{d \ln M_A(\phi)}{d \ln \phi} \right )_0 \,,
\nonumber \\
s'_A &\equiv& \left ( \frac{d^2 \ln M_A(\phi)}{d (\ln \phi)^2} \right )_0 \,,
\nonumber \\
s''_A &\equiv& \left ( \frac{d^3 \ln M_A(\phi)}{d (\ln \phi)^3} \right )_0 \,.
\end{eqnarray}
Note that the definition of $s'_A$ used in \cite{tegp} and \cite{alsing} has the opposite sign from our definition.
Recalling that $\phi = \phi_0 (1 + \Psi)$we can write
\begin{eqnarray}
M_A(\phi) &=&
 m_{A}\left [1  +  s_A \Psi + \frac{1}{2}  (s_A^2 +s'_A   -s_A) \Psi^2 
 \right .
 \nonumber \\
&&\left .+ \frac{1}{6}  (s''_A + 3 s'_A s_A - 3s'_A + s_A^3 - 3s_A^2 + 2s_A)\Psi^3 + O(\Psi^4)  \right ] 
\nonumber \\
&\equiv& m_{A} \left [1 +  {\cal S}(s_A; \Psi) \right ] \,,
\end{eqnarray}
where we define the constant mass for each body $m_A \equiv M_{A0}$.

In general relativity, neglecting pressure, the stress energy tensor can be written as (PW II, Eq.\ (2.12))
\begin{equation}
T^{\mu\nu} = \rho^* (-g)^{-1/2} u^\mu u^\nu / u^0 \,,
\end{equation}
where $\rho^*$ is identified as the ``baryonic'', or ``conserved'' mass density, $\rho^* = mn\sqrt{-g} \,u^0$, where $n$ is the number density of baryons, and $m$ is the rest mass per baryon.  It satisfies an exact continuity equation $\partial \rho^*/\partial t + \nabla \cdot (\rho^* {\bf v} ) = 0$, and implies that the baryonic mass of any isolated body is constant.   Here we identify the ``baryons'' as our compact point masses with constant mass $m_{A}$, so that 
\begin{equation}
\rho^* = \sum_A m_{A} \delta^3 ({\bf x} - {\bf x}_A) \,,
\label{rhostar}
\end{equation}
Thus, we can rewrite Eq.\ (\ref{Tmunu}) in the form
\begin{equation}
T^{\mu\nu} = \rho^* (-g)^{-1/2} u^0 v^\mu v^\nu  \left [1 +  {\cal S}(s; \Psi) \right ] \,,
\end{equation}
where $\rho^*$ is given by Eq.\ (\ref{rhostar}), and where we have substituted $u^\mu = u^0 v^\mu$, with $v^\mu = dx^\mu/dt =  (1, {\bf v})$ being the ordinary velocity.  We have dropped the subscript from the variable $s$ in $\cal S$ because it will be assigned a label $A$ wherever the delta function that is implicit in $\rho^*$ corresponds to body $A$.   Thus, we arrive at a conversion from the $\sigma$-densities of Eq.\ (\ref{sigmas}) to $\rho^*$, given by
\begin{eqnarray}
\sigma &=& \rho^*  (-g)^{-1/2} u^0 (1+v^2)  \left [1 +  {\cal S}(s; \Psi) \right ]
 \,,
\nonumber \\
\sigma^i &=&  \rho^*  (-g)^{-1/2} u^0 v^i  \left [1 +  {\cal S}(s; \Psi) \right ]
 \,,
\nonumber \\
\sigma^{ij} &=&  \rho^*  (-g)^{-1/2} u^0 v^i v^j  \left [1 +  {\cal S}(s; \Psi) \right ]
 \,.
\end{eqnarray} 
To convert $\sigma_s$, recall that 
\begin{eqnarray}
T &=& g_{\mu\nu} T^{\mu\nu}  
\nonumber \\
&=&
-\rho^* (-g)^{-1/2} (u^0)^{-1}  \left [1 +  {\cal S}(s; \Psi) \right ] \,,
\end{eqnarray}
and that $\varphi = 1 +  \Psi$, $\partial/\partial \varphi =  \partial/\partial \Psi$.  Consequently
\begin{eqnarray}
\sigma_s &=& - T + 2\varphi \frac{\partial T}{\partial \varphi} 
\nonumber \\
&=& \rho^* (-g)^{-1/2} (u^0)^{-1} 
\left [ 1 +  {\cal S} - 2(1+ \Psi) \frac{\partial {\cal S}}{\partial \Psi} \right ]  
\nonumber \\
&=& \rho^* (-g)^{-1/2} (u^0)^{-1} \bigl [ (1-2s) + {\cal S}_s (s; \Psi) \bigr ] \,.
\end{eqnarray}
Defining
\begin{eqnarray}
a_s &\equiv& s^2 +s'   - \frac{1}{2} s  \,,
\nonumber \\
{a_s}' &\equiv& s'' + 2ss' - \frac{1}{2} s'  \,,
\nonumber \\
b_s &\equiv& {a_s}' - a_s + sa_s \,,
\end{eqnarray}
we can write
\begin{eqnarray}
{\cal S} (s; \Psi) &=& s \Psi + \frac{1}{4} (2a_s -s) \Psi^2 
+ O(\Psi^3) \,,
\nonumber \\
{\cal S}_s (s; \Psi) &=&  - 2a_s \Psi - b_s \Psi^2 
+ O(\Psi^3) \,.
\end{eqnarray}

Substituting the expansion for the metric, Eq.\ (\ref{metricexpand}), and for the metric potentials, Eq.\ (\ref{expandNKB}), we obtain to the 2.5PN order required for the equations of motion,
\begin{subequations}
\begin{eqnarray}
\sigma &=& \rho^*\left[1+\epsilon \left(\frac{3}{2} v^2 -G(1-\zeta) U_\sigma + G\zeta (5+2s) U_{s \sigma} \right) 
+\epsilon^2  \left(  \frac{7}{8} v^4 
+\frac{5}{2} G^2 (1-\zeta)^2 U_\sigma^2  \right. \right.
\nonumber
\\
&& 
\left. \left .+ \frac{1}{2} G(1-\zeta) v^2 U_\sigma -4 G (1-\zeta) v^i V_\sigma^i +\frac{3}{2} (5+2s) G\zeta v^2 U_{s \sigma}
-  (5 + 2s) G^2 \zeta (1-\zeta)U_\sigma
U_{s \sigma} 
\right. \right.
\nonumber
\\
&& \left. \left . 
+\frac{1}{2} (15 + 18s + 4 a_s) G^2 \zeta^2 U_{s \sigma}^2
+\frac{3}{4} B_1-\frac{1}{4} N_1 
+\frac{1}{2} (5+ 2s)  \Psi_1 \right) 
\right .
\nonumber
\\
&&
\left. + \epsilon^{5/2}\left(2 N_{1.5}+\frac{1}{2} (5+ 2s)  \Psi_{1.5}\right) 
+{O}(\epsilon^3)  \right]  \,,
\label{sigma0PN}
\\
\sigma^i&=&\rho^*v^i\left[1+\epsilon \biggl( \frac{1}{2} v^2 
- G(1-\zeta)U_\sigma
+G \zeta (5+2s) U_{s \sigma} \biggr) 
+{O}(\epsilon^2) \right]  \,,
\label{sigmaiPN}
\\
\sigma^{ij}&=&\rho^* v^i v^j \left [1 +{O}(\epsilon) \right ] \,,
\label{sigmaijPN}
\\
\sigma^{ii}&=&\rho^* v^2 \left[1+\epsilon\biggl( \frac{1}{2}v^2
-G(1-\zeta)U_\sigma
+G \zeta (5+2s) U_{s \sigma}\biggr)+{O}(\epsilon^2) \right] \,,
\label{sigmaiiPN}
\\
\sigma_s &=& \rho^*\left [ (1-2s) - \epsilon  \left \{ \frac{1}{2}  (1-2s) v^2 + 3 G (1-\zeta)  (1-2s) U_\sigma 
- 3 G \zeta \left ( 1-2s - \frac{4}{3} a_s \right ) U_{s \sigma} \right \}
\right .
\nonumber
\\
&& \left .+\epsilon^2\biggr\{ 
-\frac{1}{8} (1-2s)v^4
+ \frac{21}{2} G^2(1-\zeta)^2 (1-2s) U_\sigma^2
-\frac{1}{2}G(1-\zeta)(1-2s)v^2 U_\sigma
\right .
\nonumber
\\
&&
\left . 
+4G(1-\zeta)(1-2s)v^iV_\sigma^i 
-\frac{3}{2} G \zeta \left((1-2s) -\frac{4}{3}a_s\right)v^2 U_{s\sigma}
-9 G^2 \zeta (1-\zeta) \left ( 1-2s - \frac{4}{3}a_s \right ) U_\sigma U_{s\sigma} 
\right .
\nonumber
\\
&&
\left .
+ \frac{3}{2} G^2 \zeta^2 \left ( 1-2s- 8a_s- \frac{8}{3} b_s\right )
U_{s\sigma}^2
+\frac{1}{4} (1-2s)B_1
-\frac{3}{4}(1-2s)N_1 
+\frac{3}{2}\left(1-2s-\frac{4}{3}a_s\right) \Psi_1\biggr\}
\right .
\nonumber
\\
&&
\left .
+\epsilon^{5/2} \biggr\{ \frac{3}{2} \left(1-2s-\frac{4}{3} a_s\right) \Psi_{1.5}\biggr\}+{O}(\epsilon^3)  \right] \,,
\label{sigmasPN}
\end{eqnarray}
\label{sigmaPN}
\end{subequations}
where $U_\sigma$, $U_{s\sigma}$ and $V^i_{\sigma}$ are defined in terms of the $\sigma$-densities.  

Substituting these formulas into the definitions of $U_\sigma$, $U_{s\sigma}$ and the other potentials defined in terms of $\sigma$, we can convert all potentials into new versions defined in terms of $\rho^*$, plus PN corrections.  For example, we find that the ``Newtonian'' potentials $U_\sigma$ and $U_{s\sigma}$ become
\begin{eqnarray}
U_\sigma&=&
U+
\epsilon \left\{
\frac{3}{2}\Phi_1-G(1-\zeta)\Phi_2
+ 6G\zeta\Phi_{2s}- G\zeta\Phi_{2s}^s\right\}
\nonumber
\\
&&
+\epsilon^2 \left\{
\frac{7}{8}\Sigma(v^4)
+ \frac{5}{2}G(1-\zeta)\Sigma(\Phi_1)
+\frac{1}{2}G(1-\zeta)\Sigma(v^2 U)
-4G(1-\zeta)\Sigma(v^i V^i)
-\frac{1}{2}G(1-\zeta)\,\Sigma(\ddot{X})
\right.
\nonumber
\\
&&
-G^2(1-\zeta)^2\Sigma(\Phi_2) 
+\frac{3}{2}G^2(1-\zeta)^2\Sigma(U^2)
+9G\zeta\Sigma(v^2 U_s)
-\frac{3}{2}G\zeta \Sigma_s(v^2 U_s)
+\frac{1}{2}G\zeta\,\Sigma_s(\Phi_1^s)
\nonumber
\\
&&
 -3G\zeta\,\Sigma(\Phi_1^s) 
+3G\zeta\,\Sigma(\ddot{X}_s)
-\frac{1}{2}G\zeta\,\Sigma_s(\ddot{X}_s)
-G^2\zeta(1+12\lambda_1+5\zeta)\Sigma(\Phi_{2s}^s)
+G^2\zeta(2\lambda_1+\zeta)\Sigma_s(\Phi_{2s}^s)
\nonumber
\\
&&
+G^2\zeta(1+17\zeta-6\lambda_1)\Sigma(U_s^2)
- \frac{1}{2}G^2\zeta(11\zeta-2\lambda_1)\Sigma_s(U_s^2)
-6G^2\zeta(1-\zeta)\Sigma(U U_s)
+G^2\zeta(1-\zeta)\Sigma_s(U U_s)
\nonumber
\\
&&
+2G^2\zeta^2\Sigma(a_s U_s^2)
-6G^2\zeta(1-\zeta)\Sigma(\Phi_2^s) 
+G^2\zeta(1-\zeta)\Sigma_s( \Phi_2^s) 
 -24 G^2 \zeta^2\,\Sigma(\Sigma(a_s U_s)) 
 +4G^2\zeta^2\,\Sigma_s(\Sigma(a_s U_s)) 
  \biggr \}
\nonumber
\\
&&+\epsilon^{5/2}\biggl\{ 
- \frac{4}{3}\stackrel{(3)\quad}{{\cal I}^{kk}(t)} U 
-\frac{1}{6}\stackrel{(3)\quad}{{\cal I}_s^{kk}(t)} (6U - U_s )
+ \frac{1}{3} \stackrel{(3)\quad}{{\cal I}_s^{j}(t)} 
\left ( 6x^j U - x^j U_s - 6X^{,j} + X_s^{,j} \right )
-\dot{M}_s(t) (6U - U_s) \biggr\}
\nonumber
\\
&&
+{O}(\epsilon^3) \,,\\
U_{s\sigma}&=&U_s
+\epsilon \biggr\{
-\frac{1}{2}\Phi_1^s
-3G(1-\zeta)\Phi_2^s
+3G\zeta\Phi_{2s}^s
-4G\zeta \Sigma(a_s U_s)
 \biggr\}
\nonumber
\\
&&+\epsilon^2 \biggl \{
-\frac{1}{8}\Sigma_s(v^4)
-\frac{1}{2}G(1-\zeta)\Sigma_s(\Phi_1)
-\frac{1}{2}G(1-\zeta)\Sigma_s(v^2 U)
+4G(1-\zeta)\Sigma_s(v^i V^i)
-\frac{3}{2}G(1-\zeta)\Sigma_s(\ddot{X})
\nonumber
\\
&&
+G^2(1-\zeta)^2\Sigma_s(\Phi_2)
+\frac{11}{2}G^2(1-\zeta)^2\Sigma_s(U^2)
-\frac{3}{2}G\zeta\Sigma_s(v^2 U_s)
+2G\zeta\Sigma(a_s v^2 U_s) 
-\frac{3}{2}G\zeta\Sigma_s(\Phi_1^s)
+2G\zeta\Sigma(a_s \Phi_1^s)
\nonumber
\\
&&
-2G\zeta\Sigma(a_s\ddot{X}_s)
+\frac{3}{2}G\zeta\Sigma_s(\ddot{X}_s)
+4G^2\zeta(2\lambda_1+\zeta)\Sigma(a_s \Phi_{2s}^s)
+G^2\zeta(1-4\zeta-6\lambda_1)\Sigma_s(\Phi_{2s}^s)
-4G^2\zeta(4\zeta-\lambda)\Sigma(a_s U_s^2) 
\nonumber
\\
&&
+ \frac{1}{2}G^2\zeta(2+7\zeta-6\lambda_1)\Sigma_s(U_s^2)
-4G^2\zeta^2\Sigma(b_s U_s^2) 
-9G^2\zeta(1-\zeta)\Sigma_s(U U_s)
+12G^2\zeta(1-\zeta)\Sigma(a_s U U_s)
\nonumber
\\
&&
-12G^2\zeta^2\Sigma_s(\Sigma(a_s U_s))
-3G^2\zeta(1-\zeta)\Sigma_s(\Phi_2^s)
+4G^2\zeta(1-\zeta)\Sigma(a_s\Phi_2^s)
+16 G^2\zeta^2\,\Sigma(a_s \Sigma(a_s U_s)) 
 \biggr \}
\nonumber
\\
&&
+\epsilon^{5/2}\left \{  
-\frac{1}{6} \stackrel{(3)\quad}{{\cal I}_s^{kk}(t)} \left (3U_s - 4  \Sigma(a_s) \right )
+ \frac{1}{3} \stackrel{(3)\quad}{{\cal I}_s^{j}(t)}  \left ( 3 x^jU_s - 4x^j \Sigma (a_s) - 3 X_s^{,j} + 4 X (a_s)^{,j} \right ) 
-\dot{M}_s(t)  \left ( 3U_s - 4 \Sigma(a_s) \right )
\right \}
\nonumber
\\
&&
+{O}(\epsilon^3) \,,
\end{eqnarray}
while the relevant PN potentials become
\begin{eqnarray}
\Phi_{1\sigma} &=& \Phi_1 
+\epsilon \left\{
\frac{1}{2} \Sigma(v^4) - G(1- \zeta) \Sigma(v^2 U) 
+ 6G\zeta \Sigma(v^2 U_s) - G \zeta \Sigma_s (v^2 U_s)
\right\}
+{O}(\epsilon^2)\,,
\\
\Phi_{2\sigma} &=& \Phi_2
+\epsilon \biggl \{
\frac{3}{2} \Sigma(v^2 U) + \frac{3}{2} \Sigma(\Phi_1)
- G(1- \zeta) \Sigma(U^2)
- G(1- \zeta) \Sigma(\Phi_2)
\nonumber 
\\
&& 
+ 6G\zeta \Sigma(UU_s) - G\zeta \Sigma_s(UU_s) 
+6G\zeta \Sigma(\Phi_{2s}) - G\zeta \Sigma(\Phi_{2s}^s)
\biggr \}
+{O}(\epsilon^2)\,,
\\
\Phi^s_{2\sigma} &=& \Phi^s_2
+\epsilon \biggl \{
-\frac{1}{2} \Sigma_s(v^2 U) + \frac{3}{2} \Sigma_s(\Phi_1)
-3 G(1- \zeta) \Sigma_s(U^2)
- G(1- \zeta) \Sigma_s(\Phi_2)
\nonumber 
\\
&& 
+ 3G\zeta \Sigma_s (UU_s)
-4 G\zeta \Sigma(a_s UU_s)
+6G\zeta \Sigma_s(\Phi_{2s}) 
- G\zeta \Sigma_s(\Phi_{2s}^s)
\biggr \}
+{O}(\epsilon^2)\,,
\\
\Phi_{2s\sigma} &=& \Phi_{2s}
+\epsilon \biggl \{
\frac{3}{2} \Sigma(v^2 U_s) - \frac{1}{2} \Sigma(\Phi_1^s)
- G(1- \zeta) \Sigma(UU_s)
- 3G(1- \zeta) \Sigma(\Phi_2^s)
\nonumber 
\\
&& 
+ 6G\zeta \Sigma(U_s^2)
- G\zeta \Sigma_s(U_s^2)
+3 G\zeta \Sigma(\Phi_{2s}^s)
-4 G\zeta \Sigma(\Sigma(a_s U_s))
\biggr \}
+{O}(\epsilon^2)\,,
\\
\Phi_{2s\sigma}^s &=& \Phi_{2s}^s
+\epsilon \biggl \{
-\frac{1}{2} \Sigma_s(v^2 U_s) - \frac{1}{2} \Sigma_s(\Phi_1^s)
- 3G(1- \zeta) \Sigma_s(UU_s)
- 3G(1- \zeta) \Sigma_s(\Phi_2^s)
\nonumber 
\\
&& 
+ 3G\zeta \Sigma_s(U_s^2)
- 4 G\zeta \Sigma(a_s U_s^2)
+3 G\zeta \Sigma_s(\Phi_{2s}^s)
-4 G\zeta \Sigma_s(\Sigma(a_s U_s))
\biggr \}
+{O}(\epsilon^2)\,,
\\
\ddot{X}_\sigma &=& \ddot{X} 
+\epsilon \biggl \{\frac{3}{2} \ddot{X}(v^2) - G(1-\zeta) \ddot{X}(U)
+ 6G \zeta \ddot{X}(U_s) - G\zeta \ddot{X}_s (U_s) 
\biggr \}
+{O}(\epsilon^2)\,,
\\
\ddot{X}_{s\sigma} &=& \ddot{X}_s 
+\epsilon \biggl \{- \frac{1}{2} \ddot{X}_s(v^2) - 3G(1-\zeta) \ddot{X}_s(U)
+ 3G \zeta \ddot{X}_s(U_s) - 4G\zeta \ddot{X}(a_s U_s) 
\biggr \}
+{O}(\epsilon^2)\,,
\\
V_\sigma^i&=&V^i+\epsilon \left\{\frac{1}{2}\Sigma(v^i v^2)-G(1-\zeta) V_2^i+6G\zeta V_{2s}^i-G\zeta \Sigma_s(v^i U_s)  \right\}
+{O}(\epsilon^2)\,,
\end{eqnarray}
where all potentials are now defined
in terms of the density $\rho^*$, and including, where needed, the sensitivity factors $s$, $a_s$ and $b_s$.  In manipulating these expressions, we have made use of the identities, valid for any function $f$,  $\Sigma(sf) = [\Sigma(f) - \Sigma_s(f)]/2$ and $\Sigma(x^i \, f) = x^i \Sigma(f) - X^{,i}(f)$.  The potentials $U$ and $U_s$ will henceforth be given by 
\begin{eqnarray}
U &=& \int_{\cal M} \frac{\rho^*(t,{\bf x}^\prime)}{|{\bf x}-{\bf x}^\prime | }d^3x^\prime \,,
\nonumber \\
U_s &=&  \int_{\cal M} \frac{\bigl (1-2s({\bf x}^\prime) \bigr ) \rho^* (t,{\bf x}^\prime)}{|{\bf x}-{\bf x}^\prime | } d^3x^\prime \,.
\end{eqnarray}
In some cases we will use the same notation as before, to avoid a proliferation of hats, tildes or subscripts.
We redefine the $\Sigma$, $X$ and $Y$ potentials by
\begin{subequations}
\begin{eqnarray}
\Sigma (f) &\equiv& \int_{\cal M} \frac{\rho^*(t,{\bf x}^\prime)f(t,{\bf x}^\prime)}{|{\bf x}-{\bf x}^\prime | } d^3x^\prime = P(4\pi\rho^* f) \,,
\\
\Sigma^i (f) &\equiv& \int_{\cal M} \frac{\rho^* (t,{\bf x}^\prime) v'^if(t,{\bf
x}^\prime)}{|{\bf x}-{\bf x}^\prime | } d^3x^\prime = P(4\pi\rho^* v^i f) \,,
\\
\Sigma^{ij} (f) &\equiv& \int_{\cal M} \frac{\rho^* (t,{\bf x}^\prime)
v'^iv'^j f(t,{\bf x}^\prime)}{|{\bf x}-{\bf x}^\prime | } d^3x^\prime 
= P(4\pi\rho^* v^iv^j f) \,,
\\
\Sigma_s (f) &\equiv& \int_{\cal M} \frac{\bigl (1-2s({\bf x}^\prime) \bigr )\rho^* (t,{\bf x}^\prime)f(t,{\bf x}^\prime)}{|{\bf x}-{\bf x}^\prime | } d^3x^\prime = P(4\pi (1-2s) \rho^* f) \,,
\\
X(f)  &\equiv& \int_{\cal M} {\rho^* (t,{\bf x}^\prime)f(t,{\bf
x}^\prime)}
{|{\bf x}-{\bf x}^\prime | } d^3x^\prime  \,,
\\
Y(f) &\equiv& \int_{\cal M} {\rho^* (t,{\bf x}^\prime)f(t,{\bf x}^\prime)}
{|{\bf x}-{\bf x}^\prime |^3 } d^3x^\prime  \,,
\end{eqnarray}
\end{subequations}
and their obvious counterparts $X^i$, $X^{ij}$, $X_s$, $Y^i$, $Y^{ij}$, $Y_s$, and so on.   With this new convention, all the potentials defined in Eqs.\ (\ref{potentiallist}) can be redefined appropriately.  

\subsection{Equation of motion in terms of potentials}

Pulling together all the potentials expressed in terms of $\rho^*$, inserting into the metric, Eq.\ (\ref{metricexpand}), calculating the Christoffel symbols, we obtain from Eq.\ (\ref{geodesiceq}) the equation of motion
\begin{equation}
dv^i/dt = a_N^i + \epsilon a_{PN}^i + \epsilon^{3/2} a_{1.5PN}^i
+ \epsilon^{2} a_{2PN}^i + \epsilon^{5/2} a_{2.5PN}^i + O(\epsilon^3) \,,
\label{dvdt25pn}
\end{equation}
where
\begin{eqnarray}
a_N^i &=& G(1-\zeta) U^{,i} + G\zeta (1-2s) U_s^{,i} \,,
\label{afluidN}
\end{eqnarray}
\begin{eqnarray}
a_{PN}^i &=& 
v^2 \left [ G(1-\zeta) U^{,i} - G\zeta (1-2s) U_s^{,i} \right ] 
- 4G(1-\zeta) v^i v^j U^{,j}
- v^i \left [ 3G(1-\zeta) \dot{U} - G\zeta (1-2s) \dot{U}_s \right ]
\nonumber \\
&&
- 4G^2 (1-\zeta )^2 UU^{,i}
- 4G^2  \zeta (1-\zeta ) (1-2s) UU_s^{,i}
- 2G^2 \zeta \left [ \lambda_1 (1-2s)  + 2\zeta s' \right ] U_s U_s^{,i}
\nonumber \\
&&
+8G(1-\zeta) v^j V^{[i,j]} + 4G(1-\zeta) \dot{V}^i
+\frac{1}{2} G(1-\zeta) \ddot{X}^{,i}
+ \frac{1}{2} G\zeta (1-2s) \ddot{X}_s^{,i}
\nonumber \\
&&
+ \frac{3}{2} G(1-\zeta) \Phi_1^{,i}  - \frac{1}{2} G\zeta (1-2s)  {\Phi^s_1}^{,i} 
- G^2(1-\zeta)^2 \Phi_2^{,i} 
- G^2  \zeta (1-\zeta ) (1-2s) {\Phi^s_2}^{,i}
\nonumber \\
&&
- G^2 \zeta \left [ 1-\zeta + (2 \lambda_1 + \zeta)(1-2s) \right ] 
{\Phi^s_{2s}}^{,i}
- 4G^2 \zeta^2 (1-2s) \Sigma^{,i}(a_s U_s) \,,
\label{afluidPN}
\end{eqnarray}
\begin{eqnarray}
a_{1.5PN}^i &=& \frac{1}{3} (1-2s) \stackrel{(3)}{{\cal I}_s^{i}} \,,
\label{afluid15PN}
\end{eqnarray}
\begin{eqnarray}
a_{2PN}^i &=&
4 G(1-\zeta)  v^i v^j v^k V^{j,k} + v^2 v^i \left [ G(1-\zeta) \dot{U} - G\zeta (1-2s) \dot{U}_s \right ]
\nonumber \\
&&
+v^i v^j  \left [ 4G^2(1-\zeta)^2 \Phi_2^{,j} + 4G^2 \zeta (1-\zeta) {\Phi^s_{2s}}^{,j}-2 G(1-\zeta) \Phi_1^{,j} -2 G(1-\zeta) \ddot{X}^{,j} \right ]
\nonumber \\
&&
+ v^j v^k \left [ 2 G(1-\zeta) \Phi_1^{jk,i} - 4 G(1-\zeta) \Phi_1^{ij,k}
+2 G^2(1-\zeta)^2 P_2^{jk,i} - 4 G^2(1-\zeta)^2 P_2^{ij,k}
\right .
\nonumber \\
&& \qquad
\left .
+ 2 G^2 \zeta (1-\zeta) P_{2s}^{jk,i} - 4 G^2 \zeta(1-\zeta) P_{2s}^{ij,k} \right ]
\nonumber \\
&&
+ v^2 \left [ -\frac{1}{2} G(1-\zeta) \Phi_1^{,i} 
+\frac{1}{2} G \zeta (1-2s) {\Phi^s_1}^{,i} 
- G^2 (1-\zeta)^2 \Phi_2^{,i}
+ G^2 \zeta (1-\zeta)(1-2s) {\Phi^s_2}^{,i}
\right .
\nonumber \\
&& \qquad
\left .
- G^2 \zeta \left [ 1-\zeta - (2 \lambda_1 + \zeta)(1-2s) \right ] 
{\Phi^s_{2s}}^{,i}
+ 2G^2 \zeta \left [ \lambda_1 (1-2s) + 2\zeta s' \right ]
U_s U_s^{,i}
 \right .
\nonumber \\
&& \qquad
\left .
+ 4 G^2 \zeta^2 (1-2s) \Sigma^{,i}(a_s U_s)
+ \frac{1}{2} G(1-\zeta) \ddot{X}^{,i}
- \frac{1}{2} G\zeta (1-2s) \ddot{X}_s^{,i}
\right ]
\nonumber \\
&&
+ v^i \left [ 
3  G^2 (1-\zeta)^2 \dot{\Phi}_2
- G^2 \zeta (1-\zeta) (1-2s) \dot{\Phi}^s_2
+ G^2 \zeta \left [ 3(1- \zeta) - (2\lambda_1 + \zeta) (1-2s) \right ]
\dot{\Phi}^s_{2s}
 \right .
\nonumber \\
&& \qquad
\left .
- 4G^2 \zeta^2 (1-2s) \dot{\Sigma}(a_s U_s)
- 2 G^2 \zeta \left [\lambda_1 (1-2s) + 2\zeta s' \right ]
U_s \dot{U}_s
-\frac{1}{2} G(1-\zeta) \dot{\Phi}_1
-\frac{1}{2} G \zeta (1-2s) \dot{\Phi}^s_1
 \right .
\nonumber \\
&& \qquad
\left .
- \frac{3}{2} G(1-\zeta) \stackrel{(3)}{X}
+\frac{1}{2} G \zeta (1-2s) \stackrel{(3)}{X}_s
+ 4 G^2 (1-\zeta)^2 V^k U^{,k}
+ 4 G^2 \zeta (1-\zeta) (1-2s)V^k U_s^{,k}
\right ]
\nonumber \\
&&
+ v^j \left [
8G^2 (1-\zeta)^2 V_2^{[i,j]}
+ 8 G^2  \zeta (1-\zeta) \Sigma_s^{,[i}(v^{j]} U_s) 
-16 G^2 (1-\zeta)^2 \Phi_2^{[i,j]}
+4 G (1-\zeta) \ddot{X}^{[i,j]}
 \right .
\nonumber \\
&& \qquad
\left .
+ 32 G^2 (1-\zeta)^2 G_7^{[i,j]}
- 8 G^2 \zeta (1-\zeta) P(\dot{U}_s U_s^{,[i})^{,j]}
- 16  G^2 (1-\zeta)^2 U V^{[i,j]}
-4 G (1-\zeta) \Sigma^{,[i}(v^{j]} v^2)
 \right .
\nonumber \\
&& \qquad
\left .
+ 8 G^2 (1-\zeta)^2 V^i U^{,j}
+ 8 G^2 \zeta (1-\zeta) (1-2s) V^j U_s^{,i}
-4 G(1-\zeta) \dot{\Phi}_1^{ij}
-4 G^2 (1-\zeta)^2 \dot{P}_2^{ij}
-4 G^2 \zeta (1-\zeta) \dot{P}_{2s}^{ij}
\right ]
\nonumber \\
&&
+ \frac{1}{24} G(1-\zeta) \stackrel{(4)}{Y^{,i}}
+ \frac{1}{24} G\zeta (1-2s) \stackrel{(4)}{Y_s^{,i}}
+2 G(1-\zeta) \stackrel{(3)}{X^{i}}
+ \frac{3}{4} G(1-\zeta) \ddot{X}_1^{,i}
- \frac{1}{4} G \zeta (1-2s) \ddot{X}_s^{,i}(v^2)
\nonumber \\
&&
+ 2 G(1-\zeta) \dot{\Sigma}(v^iv^2)
+ \frac{7}{8} G(1-\zeta) \Sigma^{,i}(v^4)
- \frac{1}{8} G \zeta (1-2s)\Sigma_s^{,i}(v^4)
+ \frac{9}{2} G^2 (1-\zeta)^2 \Sigma^{,i}(v^2 U)
\nonumber \\
&&
- \frac{1}{2} G^2 \zeta \left [ 3(1-\zeta) -(2\lambda_1 + \zeta)(1-2s)
\right ] \Sigma_s^{,i}(v^2 U_s)
- \frac{3}{2} G^2 \zeta (1-\zeta) (1-2s) \Sigma_s^{,i}(v^2 U)
\nonumber \\
&&
+ 2  G^2 \zeta^2 (1-2s)  \Sigma^{,i}(v^2 a_s U_s)
-4 G^2 (1-\zeta)^2 \Sigma^{,i}(v^j V^j)
+ 4 G^2 \zeta (1-\zeta) (1-2s) \Sigma_s^{,i}(v^j V^j)
- \frac{3}{2} G^2 (1-\zeta)^2 \Sigma^{,i}(\Phi_1)
\nonumber \\
&&
- \frac{3}{2} G^2 \zeta (1-\zeta) (1-2s) \Sigma_s^{,i}(\Phi_1)
+ 2 G^2 \zeta^2 (1-2s) \Sigma^{,i}(a_s \Phi^s_1)
+ \frac{1}{2} G^2 \zeta \left [ 1- \zeta + (2 \lambda_1 + \zeta)(1-2s) \right ]  \Sigma_s^{,i}(\Phi^s_1)
\nonumber \\
&&
- 6 G^2 (1-\zeta)^2 U \Phi_1^{,i}
+ 2 G^2 \zeta (1-\zeta) (1-2s) U {\Phi^s_1}^{,i}
+ G^2 \zeta \left [ \lambda_1 (1-2s) + 2 \zeta s' \right ]
U_s {\Phi^s_1}^{,i}
\nonumber \\
&&
-2 G^2 (1-\zeta)^2 \Phi_1U^{,i} 
- 2 G^2 \zeta (1-\zeta) (1-2s) \Phi_1 U_s^{,i}
+ G^2 \zeta \left [ \lambda_1 (1-2s) + 2 \zeta s' \right ]
\Phi^s_1 U_s^{,i}
\nonumber \\
&&
- 4 G^2 (1-\zeta)^2 \Phi_1^{ij}U^{,j} 
- 4 G^2 \zeta (1-\zeta) (1-2s)\Phi_1^{ij}U_s^{,j}
+ 8G^2 (1-\zeta)^2  V^j V^{j,i}
\nonumber \\
&&
+ 4 G^2 (1-\zeta)^2  V^i \dot{U}
- 4 G^2 \zeta (1-\zeta) (1-2s)V^i \dot{U}_s
\nonumber \\
&&
-2 G^2 (1-\zeta)^2 U \ddot{X}^{,i}
-2 G^2 \zeta (1-\zeta) (1-2s) U \ddot{X}_s^{,i}
- G^2 \zeta \left [ \lambda_1 (1-2s) + 2 \zeta s' \right ]
U_s \ddot{X}_s^{,i}
\nonumber \\
&&
-2 G^2 (1-\zeta)^2 \ddot{X} U^{,i}
-2 G^2 \zeta (1-\zeta) (1-2s)  \ddot{X} U_s^{,i}
- G^2 \zeta \left [ \lambda_1 (1-2s) + 2\zeta s' \right ]
\ddot{X}_s U_s^{,i}
\nonumber \\
&&
-8 G^2 (1-\zeta)^2 U \dot{V}^{i}
-\frac{1}{2} G^2 (1-\zeta)^2 \Sigma^{,i}(\ddot{X})
-\frac{1}{2} G^2 \zeta (1-\zeta) (1-2s) \Sigma_s^{,i}(\ddot{X})
\nonumber \\
&&
- \frac{1}{2} G^2 \zeta \left [ 1- \zeta + (2 \lambda_1 + \zeta)(1-2s) \right ]  \Sigma_s^{,i}(\ddot{X}_s)
-2 G^2 \zeta^2 (1-2s) \Sigma^{,i}(a_s \ddot{X}_s)
\nonumber \\
&&
-\frac{1}{2} G^2 (1-\zeta)^2 \ddot{X}_2^{,i}
- 2 G^2 \zeta^2 (1-2s) \ddot{X}^{,i}(a_s U_s)
-\frac{1}{2} G^2 \zeta (1-\zeta) (1-2s)  \ddot{X}_s^{,i}(U)
\nonumber \\
&&
- \frac{1}{2} G^2 \zeta \left [ 1- \zeta + (2 \lambda_1 + \zeta)(1-2s) \right ]  \ddot{X}_s^{,i}(U_s)
+ 4 G^2 (1-\zeta)^2  \dot{V}_2^i
- 4 G^2 \zeta (1-\zeta) \dot{\Sigma}_{s}(v^i U_s)
\nonumber \\
&&
- 8G^2 (1-\zeta)^2  \dot{\Phi}_2^i
- 6 G^2 (1-\zeta)^2 G_1^{,i}
+2 G^2 \zeta (1-\zeta) G_{1s}^{,i}
-4 G^2 (1-\zeta)^2 G_2^{,i}
-4 G^2 \zeta (1-\zeta) (1-2s) G_{2s}^{,i}
\nonumber \\
&&
+8 G^2 (1-\zeta)^2 G_3^{,i}
+8 G^2 \zeta (1-\zeta) (1-2s) G_{3s}^{,i}
+8 G^2 (1-\zeta)^2 G_4^{,i}
\nonumber \\
&&
-4 G^2 (1-\zeta)^2 G_6^{,i}
-4 G^2 \zeta (1-\zeta) (1-2s) G_{6s}^{,i}
+16 G^2 (1-\zeta)^2 \dot{G}_7^{i}
- 4 G^2 \zeta (1-\zeta) \dot{P}(\dot{U}_s U_s^{,i})
\nonumber \\
&&
+ 4 G^3 (1-\zeta)^3 U \Phi_2^{,i}
+ 4G^3 \zeta (1-\zeta) \left [ 1-\zeta + (2\lambda_1+\zeta)(1-2s) \right ] U {\Phi^s_{2s}}^{,i}
+ 4G^3 \zeta (1-\zeta)^2 (1-2s) U {\Phi^s_{2}}^{,i}
\nonumber \\
&& 
+ 2 G^3 \zeta (1-\zeta) \left [ \lambda_1 (1-2s) + 2\zeta s' \right ] U_s {\Phi^s_{2}}^{,i}
+ 2 G^3 \zeta (2\lambda_1+\zeta)\left [ \lambda_1 (1-2s) + 2\zeta s' \right ] U_s {\Phi^s_{2s}}^{,i}
\nonumber \\
&&
+ 16 G^3 \zeta^2 (1-\zeta) (1-2s) U \Sigma^{,i}(a_s U_s)
+ 8 G^3 \zeta^2  \left [ \lambda_1 (1-2s) + 2\zeta s' \right ] U_s \Sigma^{,i}(a_s U_s)
\nonumber \\
&&
+ 4 G^3 (1-\zeta)^3 \Phi_2 U^{,i}
+ 4 G^3 \zeta (1-\zeta)^2 \Phi^s_{2s} U^{,i}
+2 G^3 \zeta (1-\zeta) \left [ \lambda_1 (1-2s) + 2\zeta s' \right ] \Phi^s_2 U_s^{,i}
\nonumber \\
&&
+ 2 G^3 \zeta \left \{ 2\zeta (1-\zeta)(1-2s) + (2\lambda_1 + \zeta)[\lambda_1  (1-2s) + 2\zeta s'] \right \} \Phi^s_{2s} U_s^{,i}
+ 4 G^3 \zeta (1-\zeta)^2 (1-2s)  \Phi_{2} U_s^{,i}
\nonumber \\
&&
+ 8 G^3 \zeta^2  \left [ \lambda_1 (1-2s) + 2\zeta s' \right ] \Sigma(a_s U_s) U_s^{,i}
\nonumber \\
&&
+ 8 G^3 (1-\zeta)^3 U^2 U^{,i}
+ 8 G^3 \zeta (1-\zeta)^2 (1-2s) U^2 U_s^{,i}
+ 8 G^3 \zeta (1-\zeta) \left [ \lambda_1 (1-
2s) + 2 \zeta s' \right ] U U_s U_s^{,i}
\nonumber \\
&&
+ G^3 \zeta \left [ (8 \lambda_1^2 -2\zeta \lambda_1 - 2\lambda_2)(1-2s) +12 \lambda_1 \zeta s' -4 \zeta^2 s'' \right ] U_s^2 U_s^{,i}
\nonumber \\
&&
- G^3 (1-\zeta)^3 \Sigma^{,i}(\Phi_2) 
- G^3  \zeta(1-\zeta)^2 \Sigma^{,i}(\Phi^s_{2s})
- G^3 \zeta(1-\zeta)^2 (1-2s)  \Sigma_s^{,i}(\Phi_2)
\nonumber \\
&&
+ G^3 \zeta \left \{ (2\lambda_1 + \zeta) \left [1-\zeta+(2\lambda_1+\zeta)(1-2s) \right] - \zeta (1-\zeta)(1-2s) \right \} \Sigma_s^{,i}(\Phi^s_{2s})
\nonumber \\
&&
+  G^3 \zeta (1-\zeta) \left [ 1-\zeta + (2\lambda_1+\zeta)(1-2s) \right ] \Sigma_s^{,i}(\Phi^s_{2})
+ 4 G^3 \zeta^2 (1-\zeta)(1-2s)  \Sigma^{,i}(a_s \Phi^s_{2})
\nonumber \\
&&
+4 G^3 \zeta^2 (2\lambda_1 + \zeta)(1-2s) \Sigma^{,i}(a_s \Phi^s_{2s})
\nonumber \\
&&
+ 16 G^3 \zeta^3 (1-2s) \Sigma^{,i}(a_s \Sigma(a_s U_s))
+ 4 G^3 \zeta^2 \left [ 1-\zeta + (2\lambda_1+\zeta)(1-2s) \right ] \Sigma_s^{,i}(\Sigma(a_s U_s))
\nonumber \\
&&
+\frac{3}{2} G^3 (1-\zeta)^3 \Sigma^{,i}(U^2)
+ \frac{3}{2} G^3 \zeta (1-\zeta)^2 (1-2s) \Sigma_s^{,i}(U^2)
+ G^3 \zeta (1-\zeta)^2 \Sigma^{,i}(U_s^2)
\nonumber \\
&&
+ \frac{1}{2} G^3 \zeta \left \{ (2\lambda_1 + \zeta)(1-\zeta)+ (1-2s)\left [\zeta(1-\zeta) + \zeta(2\lambda_1+1)+16 \lambda_1^2 - 4\lambda_2 \right ]\right \} \Sigma_s^{,i}(U_s^2)
\nonumber \\
&&
+ G^3 \zeta (1-\zeta) \left [ 1-\zeta + (2\lambda_1+\zeta)(1-2s) \right ] \Sigma_s^{,i}(U_s U)
+ 2 G^3 \zeta^2 \left [1-\zeta + 6\lambda_1 (1-2s) \right ]\Sigma^{,i}(a_s U_s^2)
\nonumber \\
&&
-4 G^3 \zeta^3 (1-2s) \Sigma^{,i}(b_s U_s^2)
+4 G^3 \zeta^2 (1-\zeta) (1-2s) \Sigma^{,i}(a_s U_s U)
\nonumber \\
&&
- 4 G^3 (1-\zeta)^3 P_2^{ij} U^{,j}
-4 G^3 \zeta (1-\zeta)^2 P_{2s}^{ij} U^{,j}
-4 G^3 \zeta (1-\zeta)^2 (1-2s) P_{2}^{ij} U_s^{,j}
-4 G^3 \zeta^2 (1-\zeta) (1-2s) P_{2s}^{ij} U_s^{,j} 
\nonumber \\
&&
- 4 G^3 (1-\zeta)^3 H^{,i}
-4 G^3 \zeta (1-\zeta)^2 H_s^{,i}
-4 G^3 \zeta (1-\zeta)^2 (1-2s) H^{s\,,i}
-4 G^3 \zeta^2 (1-\zeta) (1-2s) {H^s_s}^{,i} \,,
\label{afluid2PN}
\end{eqnarray}
\begin{eqnarray}
a^i_{2.5PN} &=&
\frac{3}{5} x^j \left ( \stackrel{(5)\,}{{\cal I}^{ij}} - \frac{1}{3} \delta^{ij}  \stackrel{(5)\,}{{\cal I}^{kk}} \right )
+ 2 v^j \stackrel{(4)}{{\cal I}^{ij}}
+2 \left [  G(1-\zeta) U^{,j} + G\zeta (1-2s) U_s^{,j} \right ] \stackrel{(3)}{{\cal I}^{ij}}
\nonumber \\
&&
+ \frac{4}{3} \left [ G(1-\zeta) U^{,i} + G\zeta (1-2s) U_s^{,i} \right ] 
\stackrel{(3)\,}{{\cal I}^{kk}}
- \left [ G(1-\zeta) X^{,ijk} + G  \zeta  (1-2s)X_s^{,ijk} \right ]
\stackrel{(3)\,}{{\cal I}^{jk}}
\nonumber \\
&&
- \frac{2}{15} \stackrel{(5)\,}{{\cal I}^{ijj}}
+ \frac{2}{3} \epsilon^{qij} \stackrel{(4)\,}{{\cal J}^{qj}} 
- \frac{1}{15} (1-2s) x^j \left ( \stackrel{(5)\,}{{\cal I}_s^{ij}} + \frac{1}{2} \delta^{ij}  \stackrel{(5)\,}{{\cal I}_s^{kk}} \right )
+ \frac{1}{15} (1-2s) \left ( x^i x^j + \frac{1}{2} r^2 \delta^{ij} \right )
 \stackrel{(5)}{{\cal I}_s^{j}}
 \nonumber \\
&&
 + \frac{1}{30} (1-2s) \stackrel{(5)\quad}{{\cal I}_s^{ijj}} 
-\frac{1}{3}v^2 (1-2s) \stackrel{(3)}{{\cal I}_s^{i}}
-\frac{4}{3} G (1-\zeta) (1-2s) U \stackrel{(3)}{{\cal I}_s^{i}}
 \nonumber \\
&&
+ \frac{1}{6} v^i (1-2s) \left ( 2x^j \stackrel{(4)}{{\cal I}_s^{j}}
- \stackrel{(4)\,}{{\cal I}_s^{kk}}
- 6 \ddot{M}_s \right )
-\frac{1}{3} (1-2s) x^i \stackrel{(3)}{M_s}
\nonumber \\
&&
- \frac{1}{6} G \biggl \{ \left [ 1-\zeta + (4 \lambda_1 + \zeta)(1-2s) + 4\zeta s' \right ] U_s^{,i} 
+4  \zeta (1-2s) \Sigma^{,i}(a_s) \biggr \}  \left ( 2x^j \stackrel{(3)}{{\cal I}_s^{j}}
- \stackrel{(3)\,}{{\cal I}_s^{kk}}
- 6 \dot{M}_s \right )
\nonumber \\
&&
- \frac{1}{3} G \biggl \{ \left [ 1-\zeta + (4 \lambda_1 + \zeta)(1-2s) +  4\zeta s' \right ]   U_s +4  \zeta (1-2s)  \Sigma(a_s) \biggr \} \stackrel{(3)}{{\cal I}_s^{i}}
\nonumber \\
&&
+ \frac{1}{3} G \biggl \{ \left [ 1-\zeta + (2 \lambda_1+ \zeta)(1-2s) \right ]
X_s^{,ij} +4\zeta (1-2s) X_s^{,ij}(a_s) \biggr \} \stackrel{(3)}{{\cal I}_s^{j}}
 \,.
\label{afluid25PN}
\end{eqnarray}
We next turn to the problem of expressing these equations explicitly in terms of positions and velocities of each body in a two-body system.

\section{Equations of motion for two compact bodies}
\label{sec:2bodyeom}

We now wish to calculate the equation of motion for a member of a compact binary system.   To do this, we integrate $\rho^* dv^i/dt$ over body 1, and substitute Eq.\ (\ref{dvdt25pn}) and then Eqs.\ (\ref{afluidN}) -- (\ref{afluid25PN}).  We follow closely the methods already detailed in~\cite{patiwill2} (hereafter referred to as PWII) for evaluating the integrals of the various potentials, and so we will not repeat those details here.   Readers should consult Sec.\ III and Appendices B, C, and D of PWII for details.   In  structural terms almost all of the potentials that appear in the 2PN terms in scalar-tensor theory also appear in general relativity, apart from the differences in the types of densities that generate the potentials, for example $U_s$ vs. $U$, $X_s$ vs. $X$, $\Phi^s_{2s}$ vs. $\Phi_2$, and so on.   The only 2PN term that does not appear in GR involves the potential $P(\dot{U}_s U_s^{,i})$, but this can be evaluated using the methods described in PWII.  

Similarly, at 2.5PN order most of the moments that appear here also appear in GR, only a few, notably the scalar monopole and dipole moments $M_s$ and ${\cal I}_s^i$ are new.  Particularly new is the appearance of a $1.5$PN order term generated by the scalar dipole moment; this, of course, is the radiation-reaction counterpart of the well-known dipole gravitational radiation prediction of scalar-tensor theories.

\subsection{Conservative $1$PN and $2$PN terms}

We begin with the conservative Newtonian, 1PN and 2PN terms.  The results are, at Newtonian and 1PN orders.
\begin{eqnarray}
a_{1\, (PN)}^{i} &=& -  \frac{G\alpha m_2}{r^2} n^i +  \frac{G\alpha m_2}{r^2} n^i \biggl \{ 
 -  (1+ \bar{\gamma}) v_1^2  - (2+ \bar{\gamma})(v_2^2 - 2{\bf v}_1 \cdot {\bf v}_2)
      + \frac{3}{2}(\nvb)^2
\nonumber \\
  && \qquad
	+ \left [4+ 2\bar{\gamma} + 2 \bar{\beta}_1 \right ] \frac{G\alpha m_2}{r} 
	+ \left[5+2\bar{\gamma} + 2 \bar{\beta}_2 \right ] \frac{G\alpha m_1}{r} 
    \biggr \}
\nonumber 
\\
&& \quad + \frac{G \alpha m_2}{r^2} (v_1 - v_2)^i \left [(4 +2\bar{\gamma})\nva -(3 + 2\bar{\gamma}) \nvb \right ] \,, 
\nonumber 
\\
a_{2\, (PN)}^{i} &=& (1 \rightleftharpoons 2) \,,
\label{1PNeom}
\end{eqnarray}
where $r \equiv |{\bf x}_1 - {\bf x}_2 |$, ${\bf n} \equiv ({\bf x}_1 - {\bf x}_2)/r$,
and where the parameters $\alpha$, $\bar{\gamma}$, and $\bar{\beta}_A$ are defined in Table \ref{tab:params}.
Note that under the interchange $(1 \rightleftharpoons 2)$, ${\bf n} \to - {\bf n}$.  At 2PN order, we find
\begin{eqnarray}
a_{1\,(2PN)}^i &=& \frac{G \alpha m_2}{r^2}n^i\bigg \{
-(2 + \bar{\gamma}) \bigl  [v_2^4
-2 v_2^2(\v1v2)
+ (\v1v2)^2
+3(\nvb)^2 (\v1v2) \bigr]
\nonumber\\
&&
+\frac{3}{2} (1+\bar{\gamma})v_1^2(\nvb)^2
+\frac{3}{2} \left (3+ \bar{\gamma} \right )v_2^2(\nvb)^2
-\frac{15}{8}(\nvb)^4
\nonumber\\
&&
\quad + \frac{G \alpha m_2}{r}\biggl (
2 (2 +\bar{\gamma}) \bigl [ v_2^2-2\v1v2 \bigr ] - 2 \bar{\beta}_1 v_1^2
+\frac{1}{2} \left ( (2+\bar{\gamma})^2 + 4\bar{\delta}_2 \right ) \bigl [(\nva)^2
-2(\nva)(\nvb) \bigr ]
\nonumber\\
&&
\qquad \quad - \frac{1}{2} \left ((6-\bar{\gamma})(2+\bar{\gamma}) +8\bar{\beta}_1 - 4\bar{\delta}_2 \right ) (\nvb)^2
\biggr )
\nonumber\\
&&
\quad +\frac{G \alpha m_1}{r} \biggl (
\frac{1}{4} \left ( 5 + 4 \bar{\beta}_2 \right )   \left [ v_2^2 - 2\v1v2 \right ]
-\frac{1}{4} \left ( 15 + 8\bar{\gamma} + 4 \bar{\beta}_2 \right ) v_1^2
\nonumber\\
&&
\qquad \quad +\frac{1}{2} \left ( 17 + 18 \bar{\gamma}+\bar{\gamma}^2 -16 \bar{\beta}_2 + 4 \bar{\delta}_1 \right ) (\nvb)^2
\nonumber\\
&&
\qquad \quad
+\frac{1}{2}  \left ( 39 +26\bar{\gamma} + \bar{\gamma}^2 -8\bar{\beta}_2 + 4 \bar{\delta}_1 \right )
\left [(\nva)^2
-2(\nva)(\nvb) \right ] \biggr )
\nonumber \\
&&
\qquad \quad
-\frac{1}{4} \frac{G^2 \alpha^2 m_1^2}{r^2} \left ( 57 +44\bar{\gamma}+ 9\bar{\gamma}^2 + 16(3+\bar{\gamma})\bar{\beta}_2 +4 \bar{\delta}_1 - 8 \bar{\chi}_2 \right )
\nonumber \\
&&
\qquad \quad
-\frac{1}{2}\frac{G^2 \alpha^2 m_1m_2}{r^2} \left ( 69 + 48\bar{\gamma}+ 8\bar{\gamma}^2 + 8(3+\bar{\gamma})\bar{\beta}_2 +2(15+4\bar{\gamma}) \bar{\beta}_1  -48\bar{\gamma}^{-1}\bar{\beta}_1 \bar{\beta}_2 \right )
\nonumber \\
&&
\qquad \quad
-\frac{1}{4} \frac{G^2 \alpha^2m_2^2}{r^2} \left ( 9(2+\bar{\gamma})^2 + 16 (2+\bar{\gamma}) \bar{\beta}_1 + 4\bar{\delta}_2
-8 \bar{\chi}_1 \right )
\bigg \}
\nonumber \\
&&
+\frac{G \alpha m_2}{r^2}(v_1^i - v_2^i)\bigg \{
2 (2+\bar{\gamma}) \left [v_2^2(\nva) + \v1v2 (\nvb - \nva)
-\frac{3}{2}(\nva)(\nvb)^2 \right ]
\nonumber \\
&&
\qquad \quad
+(1+\bar{\gamma})v_1^2(\nvb)
-(5+3\bar{\gamma}) v_2^2(\nvb)
+\frac{3}{2} (3+2\bar{\gamma}) (\nvb)^3
\nonumber \\
&&
\qquad \quad
+\frac{G \alpha m_1}{4r}\bigg( \left ( 55 + 40\bar{\gamma} + 2\bar{\gamma}^2 - 16 \bar{\beta}_2 + 8 \bar{\delta} _1 \right )\nvb
- \left (63 + 40 \bar{\gamma}+ 2\bar{\gamma}^2 -8\bar{\beta}_2 
+ 8\bar{\delta} _1 \right ) \nva \bigg)
\nonumber \\
&&
\qquad \quad
-\frac{1}{2} \frac{ G \alpha m_2}{r}\bigg( \left ((2+\bar{\gamma})^2 + 4\bar{\delta}_2 \right ) \nva
+ \left (4-\bar{\gamma}^2 + 4\bar{\beta}_1 - 4\bar{\delta}_2 \right ) \nvb\bigg)
\bigg \} \,,
\nonumber \\
a_{2\, (2PN)}^{i} &=& (1 \rightleftharpoons 2) \,,
\label{2PNeom}
\end{eqnarray}
where $\bar{\delta}_A$ and $\bar{\chi}_A$ are defined in Table \ref{tab:params}. 

It is straightforward to show that these equations of motion can be derived from a two-body Lagrangian, given by
\begin{eqnarray}
L &=& - m_1\left (1 - \frac{1}{2} v_1^2 - \frac{1}{8} v_1^4 - \frac{1}{16} v_1^6 \right )
+ \frac{1}{2} \frac{G \alpha m_1 m_2}{r} 
\nonumber \\
&&
+ \frac{G \alpha m_1 m_2}{r} \left \{ \frac{1}{2} (3+2\bar{\gamma}) v_1^2 - \frac{1}{4} (7+4\bar{\gamma}) \v1v2 -\frac{1}{4} (\nva)( \nvb)  - \frac{1}{2} (1 + 2 \bar{\beta}_2) \frac{G \alpha m_1}{r} \right \} 
\nonumber \\
&&
+ \frac{G \alpha m_1 m_2}{r} \biggl \{ 
\frac{1}{8} (7+4\bar{\gamma})  \left [ v_1^4 - v_1^2 (\nvb)^2 \right ]
-(2+\bar{\gamma}) v_1^2 (\v1v2) + \frac{1}{8} (\v1v2)^2 
\nonumber \\
&&
\qquad + \frac{1}{16} (15 + 8\bar{\gamma}) v_1^2 v_2^2 + \frac{3}{16} (\nva)^2 (\nvb)^2
+ \frac{1}{4} (3+2\bar{\gamma}) \v1v2 (\nva)(\nvb) 
\nonumber \\
&&
\qquad
+ \frac{G \alpha m_1}{r} \biggl [ \frac{1}{8} \left (2+12\bar{\gamma}+7\bar{\gamma}^2+8\bar{\beta}_2 - 4\bar{\delta}_1 \right ) v_1^2 
+ \frac{1}{8} \left ( 14 + 20\bar{\gamma} + 7\bar{\gamma}^2 +4 \bar{\beta}_2 - 4 \bar{\delta}_1 \right ) v_2^2
\nonumber \\
&&
\qquad \quad
-\frac{1}{4} \left ( 7 +16\bar{\gamma}+ 7\bar{\gamma}^2 + 4 \bar{\beta}_2 - 4 \bar{\delta}_1 \right ) \v1v2
- \frac{1}{4} \left ( 14+12\bar{\gamma} + \bar{\gamma}^2 - 8 \bar{\beta}_2 + 4 \bar{\delta}_1 \right ) (\nva)(\nvb)
\nonumber \\
&&
\qquad \quad
+ \frac{1}{8} \left (28+ 20\bar{\gamma}+\bar{\gamma}^2 - 8 \bar{\beta}_2 + 4 \bar{\delta}_1 \right ) (\nva)^2
+ \frac{1}{8} \left (4 +4\bar{\gamma} +\bar{\gamma}^2  + 4 \bar{\delta}_1 \right ) (\nvb)^2 \biggr ]
\nonumber \\
&&
\qquad
+ \frac{1}{2} \frac{G^2 \alpha^2 m_1^2}{r^2} \left [ 1+ \frac{2}{3} \bar{\gamma}+ \frac{1}{6} \bar{\gamma}^2 + 2 \bar{\beta}_2 + \frac{2}{3} \bar{\delta}_1 - \frac{4}{3} \bar{\chi}_2 \right ]
\nonumber \\
&&
\qquad
+ \frac{1}{8} \frac{G^2 \alpha^2 m_1 m_2}{r^2} \left [ 19 +8\bar{\gamma}  + 8 \bar{\beta}_1+ 8 \bar{\beta}_2 - 32 \bar{\gamma}^{-1} \bar{\beta}_1 \bar{\beta}_2 \right ] \biggr \}
\nonumber \\
&&
- \frac{1}{8} G \alpha m_1 m_2 \biggl [ 2(7+4\bar{\gamma}) {\bf a}_1 \cdot {\bf v}_2 (\nvb) + {\bf n} \cdot {\bf  a}_1 (\nvb)^2 - (7+4\bar{\gamma}) {\bf n} \cdot {\bf a}_1 v_2^2 \biggr ] 
\nonumber \\
&&
+ (1 \rightleftharpoons 2) \,.
\label{lagrangian}
\end{eqnarray}
As in general relativity, the Lagrangian contains acceleration-dependent terms at $2$PN order, and thus the Euler-Lagrange equations are $(d^2/dt^2)(\delta L/\delta a^i) - (d/dt)(\delta L/\delta v^i) + \delta L/\delta x^i = 0$.
The equations of motion (absent radiation-reaction terms) admit the usual conserved quantities. The energy is given to 2PN order by
\begin{eqnarray}
E &=& m_1\left (\frac{1}{2} v_1^2 + \frac{3}{8} v_1^4 + \frac{5}{16} v_1^6 \right )
- \frac{1}{2} \frac{G \alpha m_1 m_2}{r} 
\nonumber \\
&&
+ \frac{G \alpha m_1 m_2}{r} \left \{ \frac{1}{2} (3+2\bar{\gamma}) v_1^2 - \frac{1}{4} (7+4\bar{\gamma}) \v1v2  -\frac{1}{4} (\nva)( \nvb)  + \frac{1}{2} (1 + 2 \bar{\beta}_2) \frac{G \alpha m_1}{r} \right \} 
\nonumber \\
&&
+ \frac{G \alpha m_1 m_2}{r} \biggl \{ 
\frac{3}{8} (7+4\bar{\gamma})  v_1^4 - \frac{1}{8} (13+8\bar{\gamma}) v_1^2 (\nvb)^2 
- \frac{1}{8} (55+28\bar{\gamma}) v_1^2 (\v1v2) + \frac{1}{8} (17+8\bar{\gamma}) (\v1v2)^2 
\nonumber \\
&&
\qquad + \frac{1}{16} (31+16\bar{\gamma}) v_1^2 v_2^2 + \frac{3}{16} (\nva)^2 (\nvb)^2
+ \frac{1}{4} (3+2\bar{\gamma}) \v1v2 (\nva)(\nvb) 
\nonumber \\
&&
\qquad
+ \frac{1}{8} (13+8\bar{\gamma}) \v1v2 (\nva)^2
- \frac{1}{8} (9+4\bar{\gamma}) v_1^2 (\nva)(\nvb) + \frac{3}{8} \nva (\nvb)^3
\nonumber \\
&&
\qquad
+ \frac{G \alpha m_1}{r} \biggl [ -\frac{1}{8} \left (12 -4\bar{\gamma} -7\bar{\gamma}^2 -8\bar{\beta}_2 +4\bar{\delta}_1 \right ) v_1^2 
+ \frac{1}{8} \left ( 14 + 20\bar{\gamma} + 7\bar{\gamma}^2 +4 \bar{\beta}_2 - 4 \bar{\delta}_1 \right ) v_2^2
\nonumber \\
&&
\qquad \quad
- \frac{1}{4}\left ( 12\bar{\gamma} + 7\bar{\gamma}^2 + 4 \bar{\beta}_2 -4 \bar{\delta}_1 \right ) \v1v2
- \frac{1}{4} \left ( 13 +12\bar{\gamma} + \bar{\gamma}^2 -8 \bar{\beta}_2 + 4 \bar{\delta}_1 \right ) (\nva)(\nvb)
\nonumber \\
&&
\qquad \quad
+ \frac{1}{8} \left (58+36\bar{\gamma}+\bar{\gamma}^2 - 8 \bar{\beta}_2 + 4\bar{\delta}_1 \right ) (\nva)^2
+ \frac{1}{8} \left (4+ 4\bar{\gamma} +\bar{\gamma}^2  + 4\bar{\delta}_1 \right ) (\nvb)^2 \biggr ]
\nonumber \\
&&
\qquad
- \frac{1}{2} \frac{G^2 \alpha^2 m_1^2}{r^2} \left [ 1 + \frac{2}{3} \bar{\gamma} + \frac{1}{6} \bar{\gamma}^2 + 2 \bar{\beta}_2 + \frac{2}{3} \bar{\delta}_1 - \frac{4}{3} \bar{\chi}_2 \right ]
\nonumber \\
&&
\qquad
- \frac{1}{8} \frac{G^2 \alpha^2 m_1 m_2}{r^2} \biggl [ 19 +8\bar{\gamma}  + 8 \bar{\beta}_1+ 8 \bar{\beta}_2 -32 \bar{\gamma}^{-1} \bar{\beta}_1 \bar{\beta}_2 \biggr ] \biggr \}
\nonumber \\
&&
+ ( 1 \rightleftharpoons 2 ) \,,
\label{Econserved}
\end{eqnarray}
while the total momentum is given by
\begin{eqnarray}
P^j &=& m_1 v_1^j 
\left ( 1 + \frac{1}{2} v_1^2 + \frac{3}{8} v_1^4  \right )
- \frac{1}{2} \frac{G \alpha m_1 m_2}{r} \left [ v_1^j + n^j (\nva) \right ]
\nonumber \\
&&
+ \frac{G \alpha m_1 m_2}{r} v_1^j 
\biggl \{ \frac{1}{8} (5+4\bar{\gamma}) v_1^2 
- \frac{1}{8} (7+4\bar{\gamma}) \left ( 2\v1v2 - v_2^2 \right ) 
 -\frac{1}{4} (\nva)( \nvb)  
\nonumber \\
&&
\qquad
+ \frac{1}{8} (13+8\bar{\gamma}) \left ( (\nva)^2 - (\nvb)^2 \right ) 
-  (3 +2\bar{\gamma} - \bar{\beta}_2) \frac{G \alpha m_1}{r}
 + \frac{1}{2} (7+4\bar{\gamma}) \frac{G \alpha m_2}{r} \biggr \} 
\nonumber \\
&&
+  \frac{G \alpha m_1 m_2}{r} n^j (\nva)
\biggl \{- \frac{1}{8} (9+4\bar{\gamma}) v_1^2 
+ \frac{1}{8} (7+4\bar{\gamma}) \left ( 2\v1v2 - v_2^2 \right ) 
\nonumber \\
&&
\qquad
+ \frac{3}{8} \left ( (\nva)^2 + (\nvb)^2 \right ) 
+  \frac{1}{4} (29 +16\bar{\gamma} ) \frac{G \alpha m_1}{r}
 - \frac{1}{4} (9 +8\bar{\gamma} - 8 \bar{\beta}_1) \frac{G \alpha m_2}{r} \biggr \}
 \nonumber \\
&&
+ ( 1 \rightleftharpoons 2 ) \,.
\label{Pconserved}
\end{eqnarray}

\subsection{Radiation-reaction terms}

At $1.5$PN order, the leading dipole radiation reaction term is given by
\begin{eqnarray}
a_{1\, (1.5PN)}^{i} &=& \frac{1}{3} (1-2s_1) \stackrel{(3)}{{\cal I}_s^{i}} \,, 
\nonumber 
\\
a_{2\, (1.5PN)}^{i} &=& \frac{1}{3} (1-2s_2) \stackrel{(3)}{{\cal I}_s^{i}} \,.
\label{15PNeom}
\end{eqnarray}
Because we will be working to $2.5$PN order, the scalar dipole moment ${\cal I}_s^i$ must be evaluated to post-Newtonian order, and when time derivatives of that moment generate an acceleration, the post-Newtonian equations of motion must be inserted.  Explicit two-body expressions for ${\cal I}_s^i$ and the other moments needed for the radiation-reaction terms are provided in an Appendix.  In addition to evaluating the direct $2.5$PN terms from Eq.\ (\ref{afluid25PN}) for two bodies, we must include the $1.5$PN contributions to the accelerations that occur in the $1$PN terms $\dot{V}^i$, $\ddot{X}^{,i}$ and $\ddot{X}_s^{,i}$ that appear in Eq.\ (\ref{afluidPN}).

At $2.5$PN order, the final two-body expressions take the form
\begin{eqnarray}
a^i_{1 \,(2.5PN)} &=&
\frac{3}{5} x_1^j \left ( \stackrel{(5)\,}{{\cal I}^{ij}} - \frac{1}{3} \delta^{ij}  \stackrel{(5)\,}{{\cal I}^{kk}} \right )
+ 2 v_1^j \stackrel{(4)\,}{{\cal I}^{ij}}
-\frac{1}{3}  \frac{G\alpha m_2}{r^2} n^i \stackrel{(3)\,}{{\cal I}^{kk}}
- 3  \frac{G\alpha m_2}{r^2} n^i n^j n^k
\stackrel{(3)\,}{{\cal I}^{jk}}
- \frac{2}{15} \stackrel{(5)\,}{{\cal I}^{ijj}}
+ \frac{2}{3} \epsilon^{qij} \stackrel{(4)\,}{{\cal J}^{qj}} 
\nonumber \\
&&
- \frac{1}{15} (1-2s_1) x_1^j \left ( \stackrel{(5)\,}{{\cal I}_s^{ij}} + \frac{1}{2} \delta^{ij}  \stackrel{(5)\,}{{\cal I}_s^{kk}} \right )
+ \frac{1}{15} (1-2s_1) \left ( x_1^i x_1^j + \frac{1}{2} r_1^2 \delta^{ij} \right )
 \stackrel{(5)}{{\cal I}_s^{j}}
  + \frac{1}{30} (1-2s_1) \stackrel{(5)\,}{{\cal I}_s^{ijj}} 
 \nonumber \\
&&
+ \frac{1}{6} v_1^i (1-2s_1) \left ( 2x_1^j \stackrel{(4)}{{\cal I}_s^{j}}
- \stackrel{(4)}{{\cal I}_s^{kk}}
- 6 \ddot{M}_s \right )
-\frac{1}{3} (1-2s_1) x_1^i \stackrel{(3)}{M_s}
-\frac{1}{3}v_1^2 (1-2s_1) \stackrel{(3)}{{\cal I}_s^{i}}
\nonumber \\
&&
+ \frac{1}{6} \frac{G\alpha m_2}{r^2} n^i \biggl \{  1-2s_2 - 4 \bar{\gamma}^{-1} \left [ (1-2s_1)\bar{\beta}_1 + (1-2s_2)\bar{\beta}_2 \right ] \biggr \}  
\left ( 2x_1^j \stackrel{(3)}{{\cal I}_s^{j}}
- \stackrel{(3)}{{\cal I}_s^{kk}}
- 6 \dot{M}_s \right )
\nonumber \\
&&
- \frac{1}{6} \frac{G\alpha m_2}{r} n^i n^j (1-2s_2)(1-8\bar{\beta}_2/\bar{\gamma} ) \stackrel{(3)}{{\cal I}_s^{j}}
-\frac{1}{6}  \frac{G\alpha m_2}{r}  (1-2s_1) (1 - 8 \bar{\beta}_1/\bar{\gamma} ) \stackrel{(3)}{{\cal I}_s^{i}} 
\nonumber \\
&&
+ \frac{1}{3} \frac{G\alpha m_2}{r} (s_1-s_2) ( 7 +4\bar{\gamma})\stackrel{(3)}{{\cal I}_s^{i}} \,,
\nonumber \\
a_{2\, (2.5PN)}^{i} &=& (1 \rightleftharpoons 2 )
 \,.
 \label{25PNeom}
\end{eqnarray}
We shall defer calculating the moments and their time derivatives explicitly until the next subsection, where we obtain the relative equation of motion.

\subsection{Relative equation of motion}

We now wish to find the equation of motion for the relative separation
${\bf x} = {\bf x}_1 - {\bf x}_2$, through $2.5$PN order.   We take the PN contributions to the equation of motion for body 1 and body 2 and calculate
$d^2 {\bf x}/dt^2 = {\bf a}_1 - {\bf a}_2$.  We must then express the individual velocities ${\bf v}_1$ and ${\bf v}_2$ that appear in post-Newtonian terms in terms of ${\bf v} \equiv {\bf v}_1-{\bf v}_2$.  Since velocity-dependent terms show up at $1$PN order, we need to find the transformation from ${\bf v}_1$ and ${\bf v}_2$ to ${\bf v}$ to $1.5$PN order so as to keep all corrections through $2.5$PN order.  To do this we make use of the momentum conservation law (\ref{Pconserved}).  But because of the contributions of dipole radiation reaction at $1.5$PN order, the momentum is not strictly conserved because of the recoil of the system in response to the radiation of linear momentum at dipole order.  By combining Eqs.\ (\ref{Pconserved}) and (\ref{15PNeom}), it is straightforward to show that the following quantity is constant through $1.5$PN order:
\begin{eqnarray}
m_1 v_1^i
\left ( 1 + \frac{1}{2} v_1^2  \right )
- \frac{1}{2} \frac{G \alpha m_1 m_2}{r} \left [ v_1^i + n^i (\nva) \right ] + \frac{1}{3} m_1(1-2s_1)  \ddot{\cal I}_s^i 
+ (1 \rightleftharpoons 2)   = C^i \,.
\end{eqnarray}
Setting $C^i = 0$ and combining this with the definition of $\bf v$, we find that 
\begin{eqnarray}
v_1^i &=& \frac{m_2}{m} v^i + \delta^i \,,
\nonumber \\
v_2^i  &=& -\frac{m_1}{m} v^i + \delta^i \,,
\label{relative1}
\end{eqnarray}
where
\begin{equation}
\delta^i =   \frac{1}{2} \eta \psi
\left [ \left (v^2 - \frac{G \alpha m}{r}
 \right ) v^i - \frac{G\alpha m}{r^2} {\dot r x^i} \right ]
- \frac{2}{3} \zeta \eta {\cal S}_{-} ({\cal S}_{+} + \psi {\cal S}_{-} )   \left (\frac{G \alpha m}{r} \right )^2 n^i + O(\epsilon^2) \,,
\label{relative2}
\end{equation}
where $m$ and $\eta$ are the total mass and reduced mass ratio, 
$\psi = \delta m /m = (m_1 - m_2)/m$, and 
\begin{eqnarray}
{\cal S}_{-}  &\equiv& - \alpha^{-1/2} (s_1 - s_2) \,,
\nonumber \\
{\cal S}_{+} &\equiv&  \alpha^{-1/2} (1 -s_1 - s_2) \,.
\end{eqnarray}

We also need to evaluate the multipole moments that appear in the radiation-reaction terms to the appropriate order, and then calculate their time derivatives, inserting the equations of motion to the appropriate order as required.  Explicit formulae for the moments are displayed in Appendix \ref{sec:moments}.  Combining all the various PN contributions consistently, we arrive finally at the relative equation of motion through $2.5$PN order, as given in Eqs.\ (\ref{eomfinal}) and (\ref{eomfinalcoeffs}).  Here we display the $2.5$PN coefficients:
\begin{eqnarray}
A_{2.5PN} &=& a_1 v^2 + a_2 \frac{G\alpha m}{r} + a_3 \dot{r}^2 \,,
\nonumber \\
B_{2.5PN} &=& b_1 v^2 + b_2 \frac{G\alpha m}{r} + b_3 \dot{r}^2 \,,
\label{25pnAB}
\end{eqnarray}
where
\begin{subequations}
\begin{eqnarray}
a_1 &=& 3 - \frac{5}{2} \bar{\gamma}  + \frac{15}{2} \bar{\beta}_+   +\frac{5}{8} \zeta {\cal S}_{-}^2 (9 + 4\bar{\gamma} -2\eta)
+ \frac{15}{8} \zeta \psi  {\cal S}_{-} {\cal S}_{+} \,,
\\
a_2 &=& \frac{17}{3}  + \frac{35}{6} \bar{\gamma} - \frac{95}{6} \bar{\beta}_+ 
- \frac{5}{24}\zeta {\cal S}_{-}^2  \left [ 135 + 56\bar{\gamma} + 8\eta + 32\bar{\beta}_+  \right ]
 +30 \zeta {\cal S}_{-} \left ( \frac{{\cal S}_{-} \bar{\beta}_+ + {\cal S}_{+} \bar{\beta}_{-}}{\bar{\gamma}} \right )
 \nonumber \\
&&  
-\frac{5}{8} \zeta \psi  {\cal S}_{-} \left ( {\cal S}_{+} - \frac{32}{3} {\cal S}_{-} \bar{\beta}_{-} +16 \frac{{\cal S}_{+} \bar{\beta}_+ + {\cal S}_{-} \bar{\beta}_{-}}{\bar{\gamma}} \right )  -40 \zeta \left (\frac{{\cal S}_{+} \bar{\beta}_+ + {\cal S}_{-} \bar{\beta}_{-}}{\bar{\gamma}} \right )^2 \,,
\\
a_3 &=& \frac{25}{8}  \left [ 2\bar{\gamma} - \zeta  {\cal S}_{-}^2 (1-2\eta)
  - 4\bar{\beta}_+ - \zeta \psi {\cal S}_{-} {\cal S}_{+}\right ]  \,, 
\\
b_1 &=& 1 - \frac{5}{6}\bar{\gamma} +\frac{5}{2} \bar{\beta}_+ 
-\frac{5}{24} \zeta {\cal S}_{-}^2 (7 + 4\bar{\gamma} -2\eta)
+ \frac{5}{8} \zeta \psi  {\cal S}_{-} {\cal S}_{+} \,,
\\
b_2 &=& 3 + \frac{5}{2} \bar{\gamma} -\frac{5}{2} \bar{\beta}_+ 
 -\frac{5}{24} \zeta   {\cal S}_{-}^2  \left [ 23 + 8\bar{\gamma} - 8\eta + 8\bar{\beta}_+  \right ]
 +\frac{10}{3} \zeta {\cal S}_{-}  \left ( \frac{{\cal S}_{-} \bar{\beta}_+ + {\cal S}_{+} \bar{\beta}_{-}}{\bar{\gamma}} \right )
  \nonumber \\
 &&
 -\frac{5}{8} \zeta \psi  {\cal S}_{-} \left ( {\cal S}_{+} - \frac{8}{3} {\cal S}_{-} \bar{\beta}_{-} + \frac{16}{3} \frac{{\cal S}_{+} \bar{\beta}_+ + {\cal S}_{-} \bar{\beta}_{-}}{\bar{\gamma}} \right ) \,,
 \\
b_3 &=& \frac{5}{8} \left [ 6\bar{\gamma} + \zeta  {\cal S}_{-}^2 (13 + 8\bar{\gamma}+2\eta)
  - 12\bar{\beta}_+  - 3 \zeta \psi {\cal S}_{-} {\cal S}_{+}  \right ]
  \,.
\end{eqnarray}
\label{25PNcoeffs}
\end{subequations}

\subsection{Energy loss rate}

We now wish to evaluate the rate of energy loss that is induced by the radiation-reaction terms in the equations of motion.  Because those equations of motion contain both $1.5$PN  as well as $2.5$PN contributions, we will have not only the normal ``quadrupole'' order contributions to the energy loss rate analogous to those that appear in general relativity, but also dipole contributions that are in principle larger by a factor of $1/v^2$.  Since the conventional ``counter'' for keeping track of contributions to the waveform and energy flux in the wave-zone denotes the GR quadrupole terms as ``Newtonian'' or $0$PN order, the dipole terms will, by this reckoning, be of $-1$PN order.  

To evaluate the energy loss correctly through ``Newtonian'' order, we first express the conserved energy in relative coordinates to $1$PN order.  Using the transformations (\ref{relative1}) and (\ref{relative2}) to $1$PN  order, we obtain
\begin{eqnarray}
E &=& \frac{1}{2} \mu v^2 - \mu \frac{G\alpha m}{r} + \frac{3}{8} \mu (1-3\eta) v^4 
\nonumber \\
&&  + \frac{1}{2} \mu \frac{G\alpha m}{r} \left [ (3 + 2\bar{\gamma} +\eta)v^2
 + \eta \dot{r}^2 \right ] +  \frac{1}{2} \mu \left ( \frac{G\alpha m}{r} \right )^2 (1+2\bar{\beta}_+ - 2\psi \bar{\beta}_{-}) \,.
\end{eqnarray}
We then calculate $dE/dt$, inserting the $1.5$PN and $2.5$PN acceleration terms into the leading term ${\bf v} \cdot {\bf a}$, and inserting only the $1.5$PN terms wherever accelerations occur in the time derivative of the $1$PN terms.  

Beginning with the leading term, and expressing the $1.5$PN acceleration in the form ${\bf a}_{1.5PN}  = (D/r^3) (3\dot{r} {\bf n} - {\bf v} )$, where
$D = 4\eta \zeta (G\alpha m)^2 {\cal S}_{-}^2/3$, we find for the $-1$PN term
$(dE/dt)_{-1PN} = \mu (D/r^3)(3\dot{r}^2 - v^2)$.   This can be simplified by exploiting the identity
\begin{equation}
\frac{d}{dt} \left ( \frac{\dot{r}}{r^2} \right ) = \frac{v^2 - 3 \dot{r}^2 +  {\bf x} \cdot {\bf a}}{r^3} \,.
\label{edotidentity}
\end{equation}
Thus $(v^2 - 3\dot{r}^2)/r^3$ can be written as the total time derivative of a quantity that can be absorbed as a $1.5$PN correction to the definition of $E$, leaving $(dE/dt)_{-1PN} = \mu (D/r^3) ({\bf x} \cdot {\bf a})$.  Inserting the Newtonian acceleration for $\bf a$, we obtain
\begin{equation}
(dE/dt)_{-1PN} = -\frac{4}{3} \zeta  \frac{\mu \eta}{r}  \left ( \frac{G\alpha m}{r} \right )^3 {\cal S}_{-}^2 \,.
\label{eq:edotdipole}
\end{equation}
This is in agreement with earlier calculations of the energy flux due to dipole gravitational radiation~\cite{eardley,tegp}.

However, since we are working to Newtonian order in the energy loss, we also need to include the $1$PN contributions to the acceleration that appears in Eq.\ (\ref{edotidentity}), yielding a contribution given by $\mu D (G\alpha m/r^4) (A_{PN} + \dot{r}^2 B_{PN})$, where $A_{PN}$ and $B_{PN}$ are given by Eqs.\ (\ref{eomfinalcoeffsPN}).   We then combine this with the other Newtonian order terms generated from $dE/dt$, leading to an expression of the general form
\begin{equation}
\frac{dE}{dt} = - \frac{8}{15} \frac{\mu \eta}{r} \left ( \frac{G \alpha m}{r} \right )^2 
\left [ p_1 \frac{G \alpha m}{r} v^2 + p_2  \frac{G \alpha m}{r} \dot{r}^2
+ p_3 v^2 \dot{r}^2
+ p_4 \left ( \frac{G \alpha m}{r} \right )^2 + p_5 v^4 
+ p_6 \dot{r}^4 \right ]
\label{edotgeneral}
\end{equation}
We now use an identity derived from the Newtonian equations of motion,
\begin{equation}
\frac{d}{dt} \left ( \frac{v^{2s}\dot{r}^p}{r^q} \right ) = \frac{v^{2s-2} \dot{r}^{p-1}}{r^{q+1}} \left( p v^4 - pv^2 \frac{G\alpha m}{r} - (p+q) v^2 \dot{r}^2 - 2s \frac{G\alpha m}{r} \dot{r}^2 \right ) \,.
\end{equation}
This is applicable at this PN order provided that the integers $s$ and $p$ are non-negative, $q \ge 2$ and $2s + p + 2q =7$.  Using the three possible cases 
$(s,\, p,\, q) = (1,\,1 ,\, 2), \, (0,\, 3,\, 2),\, (0,\,1,\,3)$,
we can freely manipulate the values of three of the six coefficients $p_i$ in Eq.\ (\ref{edotgeneral}).   The idea is to combine terms on the right-hand-side of Eq.\ (\ref{edotgeneral}) into a total time derivative, to move that to the left-hand-side and then to absorb it into a meaningless redefinition of $E$ (see for example, \cite{balacliff1,balacliff2} for discussion).  Thus one can easily arrange for $p_4$, $p_ 5$ and $p_6$ to vanish.   It then turns out that the coefficient $p_3$ of the term proportional to $v^2 \dot{r}^2$ is proportional to the combination of the $2.5PN$ equation-of-motion coefficients $5a_1 + 3a_3 - 15b_1 -5b_3$.  An inspection of Eqs.\ (\ref{25PNcoeffs}) reveals that this combination miraculously vanishes.   Pulling everything together, we obtain the final expression for the energy loss rate,
\begin{equation}
(dE/dt)_{0PN} = - \frac{8}{15} \frac{\mu \eta}{r} \left ( \frac{G \alpha m}{r} \right )^3 \left (\kappa_1 v^2  - \kappa_2 \dot{r}^2 \right ) \,,
\end{equation}
where
\begin{eqnarray}
\kappa_1 &=& 12 + 5\bar{\gamma} - 5\zeta {\cal S}_{-}^2 (3 + \bar{\gamma} + 2\bar{\beta}_+ )
+10 \zeta {\cal S}_{-} \left ( \frac{{\cal S}_{-} \bar{\beta}_+ + {\cal S}_{+} \bar{\beta}_{-}}{\bar{\gamma}} \right )
\nonumber \\
&& 
 +10 \zeta \psi   {\cal S}_{-}^2 \bar{\beta}_{-} - 10\zeta \psi  {\cal S}_{-} \left (  \frac{{\cal S}_{+} \bar{\beta}_+ + {\cal S}_{-} \bar{\beta}_{-}}{\bar{\gamma}} \right ) \,,
\nonumber \\
\kappa_2 &=& 11 + \frac{45}{4} \bar{\gamma} - 40 \bar{\beta}_+
- 5\zeta {\cal S}_{-}^2  \left [ 17+ 6\bar{\gamma} + \eta + 8\bar{\beta}_+  \right ]
 +90 \zeta {\cal S}_{-} \left ( \frac{{\cal S}_{-} \bar{\beta}_+ + {\cal S}_{+} \bar{\beta}_{-}}{\bar{\gamma}} \right )
 \nonumber \\
&&  
+ 40 \zeta \psi {\cal S}_{-}^2 \bar{\beta}_{-} - 30\zeta \psi  {\cal S}_{-} \left ( \frac{{\cal S}_{+} \bar{\beta}_+ + {\cal S}_{-} \bar{\beta}_{-}}{\bar{\gamma}} \right )  
-120 \zeta \left (\frac{{\cal S}_{+} \bar{\beta}_+ + {\cal S}_{-} \bar{\beta}_{-}}{\bar{\gamma}} \right )^2 \,.
\end{eqnarray}
These results are in complete agreement with the total energy flux to $-1PN$ and $0PN$ orders, as calculated by Damour and Esposito-Far\`ese~\cite{DamourEsposito92}.  (We are grateful to Michael Horbatsch for his invaluable help in verifying this agreement.)

\section{Discussion}
\label{sec:discussion}

We have used the DIRE approach based on post-Minkowskian theory to derive the explicit equations of motion in a general class of massless scalar-tensor theories of gravity for compact binary systems through $2.5$PN order.   Here we discuss the results, and compare our work with related work on scalar-tensor gravity and equations of motion.   

\subsection{General remarks and comparison with other results}
\label{sec:genremarks}

We begin by noting that, not surprisingly, the expressions are considerably more complicated than the corresponding general relativistic expressions.  Given that the results depend on the function $\omega(\phi)$ and its first and second derivatives, on the masses of each body, and on the sensitivities of each body and their derivatives, it is somewhat remarkable that the final equations of motion depend on a rather small number of parameters, as shown in the right-hand column of Table \ref{tab:params}.   The parameter $\alpha$ combines with $G$ to yield an effective two-body Newtonian coupling constant. It is not a universal constant, as it depends symmetrically on the sensitivities of each body.   The parameter $\bar{\gamma}$ and the body-dependent parameter $\bar{\beta}_A$ govern the post-Newtonian corrections, while the body-dependent parameters $\bar{\delta}_A$ and $\bar{\chi}_A$ govern the $2$PN corrections.   In the radiation-reaction terms, the sensitivities $s_A$ occur explicitly along with  $\bar{\gamma}$ and $\bar{\beta}_A$.  

The relative simplicity of the parameters at $1$PN and $2$PN orders has been noted before.  Damour and Esposito-Far\`ese~\cite{DamourEsposito92,DamourEsposito96} (DEF hereafter) studied a class of multi-scalar-tensor theories, but worked in the Einstein representation, where the gravitational action was pure general relativity, augmented by a free action for the scalar fields.  This is a non-metric representation of the theory, since the scalar field(s) couple to normal matter via a function $A(\varphi)$ (here we will focus on a single scalar field).  For a compact body with mass $\tilde{m}(\varphi)$ (using the Eardley ansatz), the effective matter action depends on the product $A(\varphi)\tilde{m}(\varphi)$.  The scalar field $\phi$ of our Jordan representation is given by $\phi = A(\varphi)^{-2}$, and $3 + 2\omega(\phi) = (d\ln A/d\varphi)^{-2}$.  Using a diagrammatic approach, DEF showed that the important quantities involved derivatives of $A(\varphi)\tilde{m}(\varphi)$ with respect to $\varphi$, and consequently (in our language) $\omega$ and $s_A$ and their derivatives always combined in specific ways, leading to relatively few parameters.  Table \ref{tab:dictionary} gives a dictionary that translates from our parameters to those of DEF for the case of two bodies.  Interestingly, our parameters $\bar{\delta}_A$ do not appear in DEF's list, so far as we could tell.  

\begin{table}
\caption{\label{tab:dictionary} Dictionary of parameters used in the equations of motion.  DEF refers to Ref. \cite{DamourEsposito92,DamourEsposito96}; TEGP refers to Sec.\ 11.3 of Ref. \cite{tegp}; PPN refers to the parametrized post-Newtonian limit of weakly gravitating bodies}
\begin{ruledtabular}
\begin{tabular}{cccc}
This paper&DEF&TEGP&PPN limit\\
\hline
$G\alpha$&$G_{12}$&${\cal G}_{12}$& 1\\
$\bar{\gamma}$&$\bar{\gamma}_{12}$&$\frac{3}{2} ({\cal B}_{12}/{\cal G}_{12} -1 )$&$ \gamma -1$\\
$\bar{\beta}_1$&$\beta^1_{22}$&$\frac{1}{2} ( {\cal D}_{122}/{\cal G}_{12}^2 -1)$&$ \beta -1 $\\
$\bar{\beta}_2$&$\beta^2_{11}$&$\frac{1}{2} ( {\cal D}_{211}/{\cal G}_{12}^2 -1)$&$ \beta -1 $\\
$\bar{\delta}_1$&$-$&$-$&$-$\\
$\bar{\delta}_2$&$-$&$-$&$-$\\
$\bar{\chi}_1$&$-\frac{1}{4}\epsilon^1_{222}$&$ -$&$- $\\
$\bar{\chi}_2$&$-\frac{1}{4}\epsilon^2_{111}$&$ -$&$- $\\
$\bar{\gamma}^{-1} \bar{\beta}_1 \bar{\beta}_2$&$-\frac{1}{2} \zeta_{1212}$&$-$&$ -$\\
\end{tabular}
\end{ruledtabular}
\end{table}

In the $1$PN limit, Will~\cite{tegp} wrote down a general $N$-body Lagrangian for compact self-gravitating bodies that could span a wide class of metric theories of gravity that embody post-Galilean invariance (so-called ``semi-conservative'' theories of gravity), and that have no ``Whitehead'' potential in the post-Newtonian limit.   Comparing our Lagrangian of scalar-tensor theory with the 2-body limit of Eq.\ (11.62) of \cite{tegp}, we can translate between our parameters and the coefficients ${\cal G}_{ab}$, ${\cal B}_{ab}$, and ${\cal D}_{abc}$ of \cite{tegp}, as shown in Table \ref{tab:dictionary}.    

The factor $1-2s_A$ appears throughout these equations.   This quantity is often called the ``scalar charge'' of the object.  From the point of view of the Einstein representation of scalar-tensor theory, it is easy to see how this factor arises.  The scalar field appears in the gravitational part of the action only in a kinetic term $g^{\mu\nu} \varphi_{,\mu} \varphi_{,\nu}$ (we assume that there is no potential $V(\varphi)$).  It does not couple to gravity other than via the metric in the kinetic term.  The effective matter action for a compact body depends on the product $A(\varphi) M(\varphi)$.  Varying this product with respect to $\varphi$ yields the quantity
\begin{equation}
A(\varphi) M(\varphi) \left ( \frac{d \ln A}{d\varphi} + \frac{d \ln M}{d \ln \phi}
\frac{d \ln \phi}{d \varphi} \right )  \delta \varphi= A(\varphi) M(\varphi) \frac{d \ln A}{d\varphi} (1 - 2s)  \delta \varphi\,,
\end{equation}
where we used the fact that $\ln \phi = -2 \ln A(\varphi)$.  Thus the factor $1-2s$ and its derivatives naturally control the source of the scalar field, as can be seen clearly in Eq.\ (\ref{sigmasPN}).   Defining a scalar charge for body $A$ in a two-body system by 
\begin{equation}
q_A \equiv \alpha^{-1/2} (1 - 2s_A) \,,
\end{equation}
we see that the quantities ${\cal S}_{\pm}$ are given by 
\begin{eqnarray}
{\cal S}_+ &=& \frac{1}{2} (q_1 + q_2 ) \,,
\nonumber \\
{\cal S}_{-} &=& \frac{1}{2} (q_1 - q_2 ) \,.
\end{eqnarray}

The scalar charge, or sensitivity of a given body depends on its internal structure.  For weakly gravitating bodies, $s \approx -\Omega/M \ll 1$, where
$\Omega \equiv -(1/2)G\int \rho^* \rho'^* |{\bf x} - {\bf x}'|^{-1} d^3x d^3x'$ is 
the Newtonian self-gravitational binding energy .  For neutron stars, values of the sensitivities range from $0.1$ to $0.3$, depending on the mass and equation of state of the body~\cite{willzaglauer,zaglauer} and can vary dramatically, depending on the specific form of $\omega(\phi)$~\cite{DamourEsposito92}.

\subsection{Weakly self-gravitating systems}

In the post-Newtonian limit with weakly self-gravitating systems, the sensitivities $s_i$ are themselves of order $\epsilon$.  If one is working purely at $1$PN order, then the effects of sensitivities in the $1$PN terms of Eq.\ (\ref{1PNeom}) will be of $2$PN order.  So the only effect of the bodies' sensitivities in this case will come from the coefficient $\alpha$ in the Newtonian term.  Consider a specific example: body $1$ with sensitivity $s_1$ resides in the field of body $2$, with sensitivity zero.  The acceleration of body $1$ is then given by
\begin{equation}
{\bf a}_1 = - \frac{Gm_2}{r^2} n^i (1 - 2\zeta s_1) \,,
\end{equation}
and thus the body's Newtonian acceleration will depend on its internal structure, a violation of the Strong Equivalence Principle, commonly known as the Nordtvedt effect.  In the PPN framework~\cite{tegp}, the Nordtvedt effect is normally expressed in terms of $\Omega$. Alternatively, since $M \approx m_0 + \Omega$, we have that $\Omega/M = d \ln M/d\ln G$.   Taking into account Eq.\ (\ref{Gdefinition}), we can connect the sensitivity $s$ to $\Omega$ by
\begin{equation}
s = \left ( \frac{d \ln M}{d \ln G} \right )_0 \left ( \frac{d \ln G}{d \ln \phi} \right )_0  =  - \frac{\Omega}{M} \left [ 1 + 4\Lambda (2+\omega_0) \right ] \,,
\label{svsomega}
\end{equation}
where $\Lambda$ is defined by
\begin{equation}
\Lambda \equiv \frac{\phi_0 (d\omega/d\phi)_0}{(4+2\omega_0)^2(3+2\omega_0)} \,.
\end{equation}
This is not the cosmological constant, but is the parameter defined in TEGP (see Eqs.\ (5.36) and (5.38)) such that the PPN parameter $\beta = 1+\Lambda$ in scalar-tensor theory (note the relationship between $\phi_0$ and $G$, which is set equal to unity in TEGP).   We also have that $\gamma = 1 - 2\zeta$.  We can then express the acceleration of body $1$ as 
\begin{equation}
{\bf a}_1 = - \frac{Gm_2}{r^2} {\bf n} \left [1 + \left ( \frac{1}{2+\omega_0} + 4\Lambda \right ) \frac{\Omega_1}{m_1}  \right ] \,.
\end{equation}
The coefficient in front of $\Omega_1/m_1$ is precisely $4\beta - \gamma -3$, as in the standard PPN framework.  

In the $1$PN terms in Eq.\ (\ref{1PNeom}), for weakly self-gravitating systems, it is easy to see from Table \ref{tab:params} that in the limit $s_i \to 0$, $\alpha \to 1$, the parameters $\bar{\gamma}$ and $\bar{\beta}_i$ tend to the PPN parameters $\gamma -1$ and $\beta -1$, respectively, as shown in Table \ref{tab:dictionary}, and thus our equations of motion at $1$PN order agree with the standard ones for ``point'' masses in scalar-tensor theory.

The radiation-reaction results can also be compared with existing work.  The $-1PN$ energy loss due to dipole gravitational radiation reaction, Eq.\ (\ref{eq:edotdipole}) is in complete agreement with calculations of the dipole energy flux~\cite{eardley,willrad,tegp}.  In comparing Eq.\ (\ref{eq:edotdipole}) with Eqs.\ (10.84) and (10.136) of \cite{tegp}, the additional factor of $[1+4\Lambda (2+\omega_0)]^2$ arises from the relation (\ref{svsomega}) between $s$ and $\Omega/M$.  

For weakly self-gravitating bodies, the Newtonian-order energy loss simplifies by virtue of setting all sensitivities equal to zero.   In this case, with $\alpha =1$,
$\bar{\gamma} = -2\zeta$,  
$\bar{\beta}_+ = \beta - 1 = \Lambda$, $\bar{\beta}_{-} =0$, ${\cal S}_{-} =0$, and ${\cal S}_+ = 1$, we obtain
\begin{eqnarray}
\kappa_1 &=& 12 - \frac{5}{2 + \omega_0} \,,
\nonumber \\
\kappa_2 &=& 11 - \frac{45}{2}\zeta  - 40 \Lambda -30 \Lambda^2/\zeta
\nonumber \\
&=& 11 - \frac{45}{8 +4\omega_0} \left [ 1 + \frac{8}{9} \left (\frac{2\Lambda}{\zeta} \right ) + \frac{1}{3} \left (\frac{2\Lambda}{\zeta} \right )^2 \right ] \,.
\end{eqnarray}
These agree completely with Eq.\ (10.136) of \cite{tegp}.

\subsection{Binary black holes}

Roger Penrose was probably the first to conjecture, in a talk at the 1970 Fifth Texas Symposium, that black holes in Brans-Dicke theory are identical to their GR counterparts~\cite{thornedykla}.  Motivated by this remark, Thorne and Dykla showed that during gravitational collapse to form a black hole, the Brans-Dicke scalar field is radiated away, in accord with Price's theorem, leaving only its constant asymptotic value, and a GR black hole~\cite{thornedykla}.  Hawking~\cite{hawking} proved on general grounds that stationary, asymptotically flat black holes in vacuum in BD are the black holes of GR.  The basic idea is that black holes in vacuum with non-singular event horizons cannot support scalar ``hair''.   Hawking's theorem was extended to the class of $f(R)$ theories that can be transformed into generalized scalar-tensor theories by Sotiriou and Faraoni~\cite{sotirioufaraoni}.   
 
For a stationary single body, it is clear from Eq.\ (\ref{sigmasPN}) that, if $s = 1/2$ and all its derivatives vanish, the only solution for the scalar field is $\phi \equiv \phi_0$, and hence the equations reduce to those of general relativity.
In the Einstein representation, this corresponds to $A(\varphi) M(\varphi) =$ constant, so that the scalar field decouples from any source, and thus must be either constant or singular.    Consequently, stationary black holes are characterized by $s = 1/2$. 

Another way to see this is to note that, because all information about the matter that formed the black hole has vanished behind the event horizon, the only scale on which the mass of the hole can depend is the Planck scale, and thus $M \propto M_{Planck} \propto G^{-1/2} \propto \phi^{1/2}$.  Hence $s = 1/2$.

If $s_A=1/2$ for each black hole in a binary system, then, as we discussed in the introduction, all the parameters $\bar{\gamma}$, $\bar{\beta}_A$, $\bar{\delta}_A$, $\bar{\chi}_A$, and ${\cal S}_\pm$ vanish identically, and $\alpha = 1-\zeta$.  But since $\alpha$ appears only in the combination with $G\alpha m_A$, a simple rescaling of each mass puts all equations into complete agreement with those of general relativity, through $2.5$PN order.  

But is $s_A = 1/2$ really true for binary black holes?  If the orbital timescale is long compared to the dynamical (quasinormal mode) timescale of each black hole, then it is plausible to assume that Hawking's theorem holds for each black hole, at least up to some PN order.  On the other hand, one could imagine a situation where each hole is distorted by the tidal forces from the companion hole, or where gravitational radiation flowing across the event horizons disrupts the stationarity needed for Hawking's theorem.  In PN language, these kinds of effects are known to be of an order higher than the $2.5$PN order achieved in this paper, so perhaps some non-GR effects might emerge at sufficiently high PN order.   Can a perturbation of the scalar field be supported sufficiently by strong gravity or by time varying fields to make any difference?  Or, without matter to support it, does any scalar perturbation get radiated away on a quasinormal-mode timescale, which is short compared to the orbital timescale, except during the merger of the two black holes?  Preliminary evidence from numerical relativity supports the latter scenario: Healy {\em et al.}~\cite{healy} introduced a very large Brans-Dicke type scalar field into the initial data of a binary black hole merger and found that, while the field affected the inspiral while it lasted, it was radiated away rather quickly, although it was not possible from the numerical data to fully quantify this.

It should be pointed out that there {\em are} ways to induce scalar hair on a black hole.  One is to introduce a potential $V(\phi)$, which, depending on its form, can help to support a non-trivial scalar field outside a black hole.   Another is to introduce matter.  A companion neutron star is an obvious choice, and such a binary system  in scalar-tensor theory is clearly different from its general relativistic counterpart (see the next subsection).   Another possibility is a distribution of cosmological matter that can support a time-varying scalar field at infinity.  This possibility has been called ``Jacobson's miracle hair-growth formula'' for black holes, based on work by Jacobson~\cite{jacobsonhair,burgess}.  Whether it is possible to incorporate such ideas into our approach is a subject for future work.

These considerations motivate us to formulate a conjecture along the following lines:  Consider a scalar-tensor theory of gravity with no potential for the scalar field, and consider two black holes with non-singular event horizons in a vacuum (no normal matter), asymptotically flat spacetime with $\phi$ at spatial infinity constant in time.  Following an initial transient period short compared to the orbital period, the orbital evolution and gravitational radiation from the binary system are identical to those predicted by GR, after a mass rescaling, independent of the initial scalar field configuration.   Aspects of this conjecture could be addressed by numerical simulations that extend the work of~\cite{healy}.
It may also be possible to address it partially by generalizing Hawking's theorem to a situation that is not strictly stationary, but yet still retains some symmetry, such as a helical Killing vector.  This will be the subject of future work.  

\subsection{Black-hole neutron-star systems}

Finally, we note the unusual circumstance that, if only one of the members of the binary system, say body 2, is a black hole, with $s_2 = 1/2$, then $\alpha = 1-\zeta$, $\bar{\gamma} = \bar{\beta}_A = 0$, and hence, through $1$PN order, the motion is again identical to that in general relativity.  This result is actually implicit in the post-Newtonian equations of motion for compact binaries in 
Brans-Dicke theory displayed in Eq.\ (11.91) of~\cite{tegp}, but was never stated explicitly there.  

At $1.5$PN order, dipole radiation reaction kicks in, since $s_1 < 1/2$.   In this case, ${\cal S}_{-} = {\cal S}_+ = \alpha^{-1/2} (1-2s_1)/2$, and thus the $1.5$PN coefficients in the relative equation of motion (\ref{eomfinal}) take the form
\begin{eqnarray}
A_{1.5PN} &=&  \frac{5}{8} Q \,,
\nonumber \\
B_{1.5PN} &=& \frac{5}{24} Q \,,
\end{eqnarray}
where
\begin{equation}
Q \equiv  \frac{\zeta}{1-\zeta} (1-2s_1)^2 = \frac{1}{3+2\omega_0} (1-2s_1)^2 \,.
\label{Qdefinition}
\end{equation}
At $2$PN order, $\bar{\chi}_A = \bar{\delta}_2 =0$, but $\bar{\delta}_1 = Q \ne 0$.   In this case, the $2$PN coefficients in (\ref{eomfinal}) take the form
\begin{eqnarray}
A_{2PN} &=& A_{2PN}^{GR} + Q \frac{G\alpha m_1}{r}
\left [ \dot{r}^2 -\frac{G\alpha m_1}{r} \right ] \,,
\nonumber \\
B_{2PN} &=& B_{2PN}^{GR} -2Q \frac{G\alpha m_1}{r} \,.
\end{eqnarray}
Finally, the $2.5$PN coefficients in Eq.\ (\ref{25pnAB}) have the form
\begin{eqnarray}
a_1 &=& 3 + \frac{5}{32} Q (9-2\eta+3\psi) \,,
\nonumber \\
a_2 &=& \frac{17}{3} - \frac{5}{96} Q (135+8\eta+ 3\psi) \,,
\nonumber \\
a_3 &=& - \frac{25}{32} Q (1-2\eta+\psi) \,,
\nonumber \\
b_1 &=& 1 - \frac{5}{96} Q (7-2\eta- 3\psi) \,,
\nonumber \\
b_2 &=& 3 - \frac{5}{96} Q (23 - 8 \eta + 3\psi) \,,
\nonumber \\
b_3 &=& \frac{5}{32}  Q (13 + 2\eta-3\psi) \,,
\end{eqnarray}
while the coefficients in the energy loss rate simplify to
\begin{eqnarray}
\kappa_1 &=& 12 - \frac{15}{4} Q \,,
\nonumber \\
\kappa_2 &=& 11 - \frac{5}{4} Q (17 + \eta) \,.
\end{eqnarray}
We  find, somewhat surprisingly, that the motion of a mixed compact binary system through $2.5$PN order differs from its general relativistic counterpart only by terms that depend on a single parameter $Q$, as defined by Eq.\ (\ref{Qdefinition}).  Furthermore, all reference to the parameters $\lambda_1$ and $\lambda_2$, related to derivatives of the coupling function $\omega(\phi)$, has disappeared, in other words, the motion of mixed compact binary systems in general scalar-tensor theories through $2.5$PN order is formally identical to that in standard Brans-Dicke theory.   The only way that a generalized scalar-tensor theory affects the motion differently than pure Brans-Dicke theory is through the value of the un-rescaled mass $m_1$ and the sensitivity $s_1$ for a neutron star of a given central density and total number of baryons.

The general conclusions reached in this paper about binary black holes and mixed binaries in scalar-tensor gravity were obtained from the near-zone gravitational fields.   If these conclusions continue to hold for the gravitational-wave signal, then gravitational-wave observations of binary black holes will be unable to distinguish between general relativity and scalar-tensor theories, and observations of mixed black-hole neutron-star binaries will be essentially unable to distinguish between general scalar-tensor theories and Brans-Dicke theory.   The  
radiative part of this problem, which will involve a derivation of the gravitational waveform to $2$PN order, together with the energy flux, will be the subject of the second paper in this series.

\acknowledgments

This work was supported in part by the National Science Foundation,
Grant Nos.\ PHY 09--65133 \& 12--60995.  We thank the Institut d'Astrophysique de Paris for
its hospitality during the completion of this work.   We are grateful to Emanuele Berti and Michael Horbatsch for critical reading of the manuscript and for helpful comments, and to Nicol\`as Yunes for useful discussions during the early part of this work.

\appendix

\section{Multipole moments for two-body systems}
\label{sec:moments}

Here we evaluate the multipole moments that appear in the radiation reaction expressions (\ref{15PNeom}) and (\ref{25PNeom}) to the order required to obtain $2.5$PN-accurate contributions.  The scalar dipole moment ${\cal I}_s^i$ in Eq.\ (\ref{15PNeom}) must be evaluated to $1$PN order.   Substituting $\tau_s$ from Eq.\ (\ref{tausPN}) and $\sigma_s$ from Eq.\ (\ref{sigmasPN}) to $1$PN order into Eq.\ (\ref{IsQ}), we obtain
\begin{equation}
{\cal I}_s^i = G \zeta m_1 x_1^i (1-2s_1) \left [ 1 - \frac{1}{2} v_1^2 - \frac{G \alpha m_2}{r} \left ( 1 - 4 \frac{\bar{\beta}_1}{\bar{\gamma}} \right ) \right ]  + ( 1 \rightleftharpoons 2 ) \,.
\end{equation}

Most of the multipole moments that appear in the $2.5$PN expressions (\ref{25PNeom}) can be evaluated to the lowest PN order, so that we may write
\begin{subequations}
\begin{eqnarray}
{\cal I}^{ij} &=& G(1-\zeta) \left (  m_1 x_1^{ij} + m_2 x_2^{ij} \right )  \,,
\\
{\cal I}^{ijk} &=& G(1-\zeta) \left (  m_1 x_1^{ijk} + m_2 x_2^{ijk} \right )  \,,
\\
{\cal J}^{qj} &=& G(1-\zeta) \epsilon^{qab} \left (  m_1 v_1^b x_1^{aj} + m_2 v_2^b x_2^{aj} \right ) \,,
\\
{\cal I}_s^{ij} &=& G\zeta  \left (  m_1 (1-2s_1) x_1^{ij} + m_2 (1-2s_2) x_2^{ij} \right )  \,,
\\
{\cal I}_s^{ijk} &=& G\zeta \left (  m_1 (1-2s_1) x_1^{ijk} + m_2 (1-2s_1) x_2^{ijk} \right )  \,.
\end{eqnarray}
\end{subequations}
The exception to this rule is the scalar monopole moment $M_s = \int_{\cal M} \tau_s d^3x$; formally it contributes at $0.5$PN order, as can be seen in Eq.\ (\ref{bigexpansiond}), but its leading contribution is constant in time, and hence it is the $1$PN correction that matters.  Inserting  $\tau_s$ and $\sigma_s$ from Eqs.\ (\ref{tausPN}) and (\ref{sigmasPN}) to $1$PN order, we obtain
\begin{equation}
M_s  = G \zeta m_1 (1-2s_1) \left [ 1 - \frac{1}{2} v_1^2 - \frac{G \alpha m_2}{r} \left ( 1 -4 \frac{\bar{\beta}_1}{\bar{\gamma}} \right ) \right ]  + ( 1 \rightleftharpoons 2 ) \,.
\end{equation}
Since the first term is constant, it can be dropped.



\end{document}